\documentclass[a4paper,11pt,oneside]{article}
\usepackage[latin1]{inputenc}
\usepackage{ngerman}
\usepackage{amsmath}
\usepackage{amssymb}
\usepackage{latexsym}
\usepackage[round]{natbib}
\usepackage{makeidx}
\usepackage{nomencl}
\usepackage{graphicx}
\usepackage[arrow, matrix, curve]{xy}
\usepackage{mathrsfs}
\usepackage{dsfont}   %%% fuer \N, \R etc
\usepackage{enumitem}
\usepackage{color}

 \newtheorem{theorem}{Theorem}[section]
 
 \newtheorem{lemma}[theorem]{Lemma}

 \newtheorem{proposition}[theorem]{Proposition}
 
 \newtheorem{assumption}[theorem]{Assumption}

 \newenvironment{proof}
   {\begin{list}{\textbf{Proof}:}
                {\setlength{\leftmargin}{0.5em}
                 \setlength{\labelwidth}{0em}
                }
   }
   {\hspace*{\fill}$\Box$\end{list}}
   
 \newenvironment{proof*}
   {\begin{list}{\textbf{Proof}:}
                {\setlength{\leftmargin}{0.5em}
                 \setlength{\labelwidth}{0em}
                }
   }
   {\end{list}}

 \newcounter{number}

 \newcounter{subnumber}

\title{Practical Tikhonov Regularized Estimators in Reproducing Kernel Hilbert Spaces
       for Statistical Inverse Problems
}
\author{Robert Hable}
\date{}

\begin{document}

\maketitle

\begin{abstract}
Regularized kernel methods such as support vector machines (SVM)
and support vector regression (SVR)
constitute a broad and flexible class of
methods which are theoretically well investigated and 
commonly used in nonparametric classification and regression problems.
As these methods are based on a Tikhonov regularization
which is also common in inverse problems,
this article investigates the use of 
regularized kernel methods for inverse problems in a unifying way.
Regularized kernel methods are based on the use of reproducing kernel Hilbert spaces
(RKHS) which lead to very good computational properties.
It is shown that similar properties remain true in solving 
statistical inverse problems and that standard software implementations
developed for ordinary regression problems can still be used for
inverse regression problems. 

Consistency of these methods and
a rate of convergence for the risk is shown under quite 
weak assumptions and
rates of convergence for the estimator are 
shown under somehow stronger assumptions.
The applicability of these methods is demonstrated in a simulation.
\end{abstract}

\section{Introduction}

One of the most important statistical inverse problems is
the inverse regression problem in which one observes 
i.i.d.\ data $(z_1,y_1),\dots,(z_n,y_n)$ from the model
\begin{eqnarray}\label{intro-heteroscedastic-model}
  Y\;=\;\big(Af_0\big)(Z)+s(Z)\varepsilon
\end{eqnarray}
in which $A$ is a (known) linear operator between suitable function spaces,
$Af_0:\mathcal{Z}\rightarrow\mathbb{R}$ is the (unknown) regression function, and
$s:\mathcal{Z}\rightarrow\mathbb{R}$ is an (unknown) scale function.
The goal is to estimate the primary function $f_0:\mathcal{X}\rightarrow\mathbb{R}$.
If $A$ has a bounded inverse $A^{-1}$, then $f_0$ can simply be estimated by
$A^{-1}\hat{g}_n$ where $\hat{g}_n$ is an ordinary estimate
of the regression function $g=Af_0$. However, in a typical inverse regression problem,
$A$ does not have a bounded inverse so that one is faced with an ill-posed
problem which has to be dealt with in different and much more complicated
ways. See, e.g., \cite{Cavalier:2011} for an overview.
Two of the most common types of estimators in inverse regression problems
are spectral cut-off estimators and Tikhonov estimators.
In case of the spectral cut-off estimator, it is assumed that
$A$ is an injective compact operator between $L_2$-spaces $L_2(\mu)$ 
and $L_2(P_{\mathcal{Z}})$
so that a singular value decomposition exits. That is, there are 
a complete orthonormal system $(v_j)_{j\in\mathbb{N}}$ of $L_2(\mu)$,
an orthonormal system $(u_j)_{j\in\mathbb{N}}$ in $L_2(P_{\mathcal{Z}})$,
and singular values 
$(\sigma_j)_{j\in\mathbb{N}}\subset(0,\infty)$
such that $Av_j=\sigma_ju_j$ and $A^\ast u_j=\sigma_j v_j$ where
$A^\ast$ denotes the adjoint operator of $A$. Then, the spectral cut-off
estimator is given by
$$\hat{f}_{n}(x)\;=\;
  \sum_{j=1}^J \frac{\hat{b}_{j,n}}{\sigma_j}v_j(x)
  \qquad\text{with}\quad \hat{b}_{j,n}=\frac{1}{n}\sum_{i=1}^n u_j(z_i)y_i
$$
where the truncation parameter $J=J_n$ acts as a regularization parameter.
This estimator is investigated in many articles on inverse regression problems, e.g.,
in \cite{Rooij:Ruymgaart:1996}, 
\cite{Mair:Ruymgaart:1996},
\cite{Bauer:Munk:2007}, 
\cite{Bissantz:Holzmann:2008}, and
\cite{Bissantz:Birke:2009}.
A disadvantage of this estimator is that one needs to know the
singular value decomposition of $A$ in order to compute the estimator
(this is also a considerable limitation for general software implementations).
Furthermore, one usually has to know the distribution $P_{\mathcal{Z}}$ of
the covariate $Z$. Most articles on spectral cut-off estimators assume that
$Z$ is uniformly distributed on $[0,1]$ or use equidistant design points. 

Methods based on Tikhonov regularizations are common in non-stochastic as well as
in statistical
settings of inverse problems. In inverse regression problems, the estimator
based on Tikhonov regularization is the minimizer
\begin{eqnarray}\label{intro-tikhonov}
  \text{arg}\min_{f\in H}
  \left(\frac{1}{n}\sum_{i=1}^n\big(y_i-\big(Af\big)(z_i)\big)^2
        +\lambda\|f\|_H^2
  \right)
\end{eqnarray}
where $H$ is a Hilbert space of functions $f:\mathcal{X}\rightarrow\mathbb{R}$.
Such an estimator has been considered, e.g., in 
\cite{OSullivan:1986},
\cite{Mathe:Pereverzev:2001},
\cite{Bissantz:Hohage:2007}, and 
\cite{Cavalier:2008}.
Most articles on Tikhonov regularization in 
statistical inverse problems
focus on rates of convergence, but simulations or applications on real data sets 
are only rarely done. One reason might be that calculating the estimator
is, in general, not an easy task and suitable software implementations
(e.g., as R-packages)
are still widely missing.
The situation gets better if reproducing kernel Hilbert spaces (RKHS) are chosen 
for $H$ in (\ref{intro-tikhonov}). These Hilbert spaces have excellent properties
from a computational point of view and, therefore, recently attract much attention
in statistics, machine learning, and approximation theory. 
Tikhonov estimators (\ref{intro-tikhonov}) in an RKHS are already used 
in the early work
\cite{Wahba:1977} and \cite{Wahba:1980}; there, a special case of model 
(\ref{intro-heteroscedastic-model}) is considered in which
the error is homoscedastic (i.e., $s\equiv 1$), $\mathcal{X}=\mathcal{Z}=[0,1]$,
the data
$z_i$, $i\in\{1,\dots,n\}$, are equidistant design points, and $A$ 
is a Fredholm integral operator of the first kind, that is,
$$Af(z)=\int K(x,z)f(x)\,\lambda(dx)
  \qquad\forall\,z\in[0,1],\;\;f\in H\,.
$$
A similar setting is also considered in \cite{Nychka:Cox:1989} which shows
rates of convergence for Tikhonov estimators (\ref{intro-tikhonov}) in an RKHS.
In ordinary nonparametric classification and regression problems,
lots of research has been done on
methods based on RKHS such as support vector machines (SVM) and
support vector regression (SVR) during the last decade. 
These methods belong to a broad class of
methods called regularized kernel methods. In an ordinary (heteroscedastic)
regression problem  
\begin{eqnarray}\label{intro-ordinary-heteroscedastic-model}
  Y\;=\;f_0(X)+s(X)\varepsilon\,,
\end{eqnarray}
the estimate of a regularized kernel method is the minimizer 
\begin{eqnarray}\label{intro-ordinary-RKM}
  \text{arg}\min_{f\in H}
  \left(\frac{1}{n}\sum_{i=1}^n L\big(x_i,y_i,f(x_i)\big)
        +\lambda\|f\|_H^2
  \right)
\end{eqnarray}
where $H$ is an RKHS and $L$ is a suitable loss function;
see, e.g., \cite{vapnik1998}, \cite{schoelkopf2002},
and \cite{steinwart2008}.
In view of (\ref{intro-tikhonov}), regularized kernel methods can also be
defined for inverse regression problems in an obvious way by
\begin{eqnarray}\label{intro-inverse-RKM}
  \text{arg}\min_{f\in H}
  \left(\frac{1}{n}\sum_{i=1}^n L\big(z_i,y_i,(Af)(z_i)\big)
        +\lambda\|f\|_H^2
  \right).
\end{eqnarray}
The goal of the present article is to investigate these methods 
(which considerably generalize the setting of the early work
in \cite{Wahba:1977}, \cite{Wahba:1980}, and \cite{Nychka:Cox:1989})
in a unifying way
in the light of the current state of research on regularized kernel methods
for ordinary regression problems (\ref{intro-ordinary-heteroscedastic-model}).
The results are not restricted to $\mathcal{X}=\mathcal{Z}=[0,1]$
but $\mathcal{X}$ may be any compact subset of $\mathcal{R}^d$ 
for any dimension $d\in\mathbb{N}$ and $\mathcal{Z}$ may be 
any Polish space. Furthermore, we also consider heteroscedastic
errors as homoscedasticity is less frequent in real data sets.
Allowing for different loss functions, on the one hand, extends the
applicability of the method from mean regression to 
tasks such as median regression, quantile regression, and classification.
On the other hand, it is well known that the least-squares loss typically
induces bad robustness properties and that regularized kernel methods
for Lipschitz-continuous loss functions (such as the absolute deviation loss)
have very good robustness properties; see 
\cite{christmannsteinwart2007},
\cite{christmann2009}, and
\cite{HableChristmann2011}.\\
Formally, using the least-squares loss simply was a computational need;
due to the nice structure of the least-squares loss,
calculating (\ref{intro-tikhonov}) is equivalent to solving the 
equality
\begin{eqnarray}\label{intro-Tikhonov-euivalent-equation}
  A^\ast Af+n\lambda f\;=\;A^\ast y\,,\qquad f\in H,
\end{eqnarray}
in $H$ and, for compact operators $A$ with singular system
$(\sigma_j;v_j,u_j)$, the solution is given by
$$\hat{f}_n\;=\;
  \sum_{j=1}^\infty \frac{\sigma_j}{\sigma_j^2+n\lambda}
     \langle y,u_n\rangle v_j\,;
$$ 
see, e.g., \cite[p.\ 117]{Engl:Hanke:1996}.
If $H$ is an RKHS, solving (\ref{intro-Tikhonov-euivalent-equation}) 
is much easier but still involves calculating the inverse of an
$n\times n$-matrix; see \cite[p.\ 654]{Wahba:1977}.
However, nowadays, enormous efforts have been made in order to
develop powerful software implementations 
for calculating regularized kernel methods 
(\ref{intro-ordinary-RKM}) such as SVM and SVR
for various loss functions. It turns out 
(Theorem \ref{theorem-empirical-representer-theorem})
that these
implementations can still be used in oder to calculate (\ref{intro-inverse-RKM}), i.e.,  
regularized kernel methods for inverse problems. 
One only has to calculate a certain ``pseudo'' kernel matrix $M$
and proceed with standard software implementations as if
$M$ was the kernel matrix $K$ from an ordinary regularized kernel method;
see Section \ref{section-setup} and 
\ref{section-simulation} for details. 

\medskip

In a non-stochastic setting, an RKHS in (\ref{intro-inverse-RKM}) 
has also been
considered in \cite{krebs:louis:2009} for the least-squares loss
and in \cite{krebs2011} for the $\varepsilon$-insensitive loss,
a common loss function in machine learning. 
Also in a non-stochastic setting,
\cite{Eggermont:LaRiccia:2012} consider an RKHS in the Tikhonov method 
(with the least-squares loss) but in a quite different way: there,
the codomain of $A$ is the subset of an RKHS while we use an RKHS
as the domain of $A$. 

\medskip

The rest of the article is organized as follows: 
Section \ref{section-setup} contains notations,
assumptions, and the general definition of
regularized kernel methods for inverse problems.
Furthermore, it is shown that the estimators uniquely exist
(Theorem \ref{theorem-unique-existence-theoretical-svm})
and that an analogue of the empirical representer theorem holds
which enables to use standard software implementations developed for
ordinary regularized kernel methods 
(Theorem \ref{theorem-empirical-representer-theorem}).
In Section \ref{section-consistency}, consistency in
the $H$-norm (which is stronger than the supremum-norm) and
a rate of convergence for the risk is shown under quite 
weak assumptions (Theorem \ref{theorem-consistency}
and Theorem \ref{theorem-rate-of-convergence-risks}).
A rate of convergence of the estimator in the $H$-norm is 
shown under somehow stronger assumptions 
(Theorem \ref{theorem-rate-of-convergence}).
Section \ref{section-simulation} contains a simulation 
in a standard example, namely the heat equation.
All proofs are deferred to the appendix;
Subsection \ref{section-appendix-addiditional-results} in the appendix
also contains a number of additional results which are needed in the proofs
of the main results and are interesting on its own.

\section{Regularized Kernel Methods in Inverse Problems}\label{section-setup}

Throughout the whole article, we deal with the general setting
summarized in the following assumption:
\begin{assumption}\label{assumption-general-setting}
  Let $P$ be a probability measure on 
  $\big(\mathcal{Z}\times\mathcal{Y},
      \mathfrak{B}_{\mathcal{Z}\times\mathcal{Y}}
   \big)
  $
  where $\mathcal{Z}$ is a Polish space, $\mathcal{Y}\subset\mathbb{R}$
  is closed and $\mathfrak{B}_{\mathcal{Z}\times\mathcal{Y}}$ is the 
  Borel-$\sigma$-algebra on $\mathcal{Z}\times\mathcal{Y}$. The marginal
  distribution of $P$ on $\mathcal{Z}$ is denoted by $P_{\mathcal{Z}}$;
  the corresponding conditional distribution on $\mathcal{Y}$ given $z$
  is denoted by $P(\cdot|z)$. That is,
  $$\int g\,dP\;=\;\iint g(z,y)\,P(dy|z)\,P_{\mathcal{Z}}(dz)
    \qquad\forall\,g\in L_1(P)\,.
  $$
  Let $\mathcal{X}\subset\mathbb{R}^d$ be compact and  
  let $k$ be a kernel on $\mathcal{X}$ 
  which is extendable
  to an $m$-times differentiable kernel on $\mathbb{R}^d$ where
  $m>\frac{d}{2}$. The RKHS of $k$ is denoted by $H$.
  The operator 
  \begin{eqnarray}\label{assumption-operator-A-continuous-on-H}
    A\;:\;\;H\;\rightarrow\;
    \mathcal{C}_{b}(\mathcal{Z})
    \qquad\text{is continuous and linear}
  \end{eqnarray}
  where $\mathcal{C}_{b}(\mathcal{Z})$ denotes the Banach space of all
  bounded, continuous functions $g:\mathcal{Z}\rightarrow\mathbb{R}$ with
  supremum-norm $\|\cdot\|_\infty$.\\
  The function $L:\mathcal{Z}\times\mathcal{Y}\times\mathbb{R}\rightarrow[0,\infty)$
  is a continuous loss function; that is, the function $t\mapsto L(z,y,t)$
  is convex for every fixed $z\in\mathcal{Z}$, $y\in\mathcal{Y}$.
\end{assumption}

In order to prove consistency and rates of convergence,
Assumption (\ref{assumption-operator-A-continuous-on-H})
will be replaced by the stronger assumption that
\begin{eqnarray}\label{assumption-operator-A-continuous-on-Cb}
  A\;:\;\;\mathcal{C}_{b}(\mathcal{X})\;\rightarrow\;
  \mathcal{C}_{b}(\mathcal{Z})
  \qquad\text{is continuous and linear}
\end{eqnarray}
later on.
However, it is always made explicit whenever 
(\ref{assumption-operator-A-continuous-on-Cb}) is assumed instead of
(\ref{assumption-operator-A-continuous-on-H}).
Note that Assumption (\ref{assumption-operator-A-continuous-on-Cb}) indeed implies
(\ref{assumption-operator-A-continuous-on-H}) because,
according to \cite[Lemma 4.23]{steinwart2008},
\begin{eqnarray}\label{sup-norm-H-norm}
  \|f\|_\infty\;\leq\;\|k\|_\infty\|f\|_H
  \qquad\forall\,f\in H.
\end{eqnarray}
Articles on inverse regression problems typically assume compactness of
the operator $A$ and it seems as if we had no such assumption here. However,
$A$ enjoys a compactness property which comes for free in this setting.  
As shown in Prop.\ \ref{prop-compactness-A}, 
it follows from (\ref{assumption-operator-A-continuous-on-Cb})
that $A:H\rightarrow\mathcal{C}_{b}(\mathcal{Z})$ is a compact operator,
but this compactness is a quite weak property.
The reason for this is that weak convergence in the RKHS of a bounded continuous
kernel is a relatively strong kind of convergence. In particular, weak convergence
in $H$ implies pointwise convergence, which is an easy consequence of the so-called
reproducing property
\begin{eqnarray}\label{reproducing-property}
  \big\langle f,\Phi(x) \big\rangle_H\;=\;f(x)
  \qquad\forall\,x\in\mathcal{X},\;\;f\in H
\end{eqnarray}
where $\Phi$ denotes the canonical feature map of $H$, i.e., $\Phi(x)=k(x,\cdot)$
for every $x\in\mathcal{X}$.

\bigskip

In most parts of the article, we will also impose the following 
assumption on the loss function $L$:
\begin{assumption}\label{assumption-on-L}
  Assume that $L$ is continuous and
  that there are $b\in\mathcal{L}_1(P)$, $c_L\in(0,\infty)$,
  and $\beta_L\in[1,2]$ such that, for every
  $z\in\mathcal{Z}$, $y\in\mathcal{Y}$, and $t\in\mathbb{R}$,
  \begin{eqnarray}\label{assumption-on-L-1}
    L(z,y,t)\;\leq\;b(z,y)+c_L |t|^{\beta_L}\;.
  \end{eqnarray}
  In addition, assume that there are $p\in[0,1]$, 
  $b_0^\prime\in\mathcal{L}_2(P_{\mathcal{Y}})$ with $b_0^\prime\geq 0$, 
  and $b_1^\prime\in[0,\infty)$
  such that, for every
  $z\in\mathcal{Z}$, $y\in\mathcal{Y}$, $a\in(0,\infty)$, and
  $t_1,t_2\in[-a,a]$,
  \begin{eqnarray}\label{assumption-on-L-2}
    \big|L(z,y,t_1)-L(z,y,t_2)\big|\;\leq\;
    \big(b_0^\prime(y)+b_1^\prime a^p\big)\cdot
    |t_1-t_2|\;.
  \end{eqnarray}
\end{assumption}
This assumption looks quite special but, indeed, covers all of the
commonly used loss functions:
least squares, hinge, truncated least squares, and logistic for 
classification; least squares, absolute distance, pinball,
epsilon-insensitive, Huber, and logistic for regression.

\medskip

For 
$D_n=\big((z_1,y_1),\dots,(z_n,y_n)\big)
  \in(\mathcal{Z}\times\mathcal{Y})^n
$ 
and $\lambda>0$,
define
\begin{eqnarray}\label{def-empirical-svm}
  f_{A,D_n,\lambda}\;=\;
  \text{arg}\min_{f\in H}
  \Bigg(\frac{1}{n}\sum_{i=1}^n L\big(z_i,y_i,(Af)(z_i)\big)
        +\lambda\|f\|_H^2
  \Bigg)
\end{eqnarray}
and the regularized empirical risk
$$\mathcal{R}_{A,D,\lambda}(f)\;=\;
  \frac{1}{n}\sum_{i=1}^n L\big(z_i,y_i,(Af)(z_i)\big)
        +\lambda\|f\|_H^2
  \qquad\forall\,f\in H\,.
$$
That is, 
$f_{A,D_n,\lambda}=\text{arg}\min_{f\in H}\mathcal{R}_{A,D,\lambda}(f)$.
The following theorem is the analogue to the well-known representer theorem in
case of ordinary regularized kernel methods (i.e. $A=\text{id}$); see, e.g., 
\cite[Theorem 5.5]{steinwart2008}. 
By use of this theorem, the optimization problem (\ref{def-empirical-svm}) in 
the infinite-dimensional function space $H$ can be reduced to a convex 
optimization problem in $\mathbb{R}^n$.
In similar but non-stochastic inverse problems, corresponding results
have already been obtained for special loss functions
in \cite[Lemma 3.1 and Theorem 3.2]{krebs:louis:2009} and \cite[Lemma 3.1]{krebs2011}.
\begin{theorem}[Empirical Representer Theorem] 
  \label{theorem-empirical-representer-theorem}
  \hfill\\
  Let Assumption \ref{assumption-general-setting} be fulfilled.
  For every
  $D_n=\big((z_1,y_1),\dots,(z_n,y_n)\big)
    \in(\mathcal{Z}\times\mathcal{Y})^n
  $ 
  and $\lambda>0$, the function $f_{A,D_n,\lambda}$
  defined by (\ref{def-empirical-svm}) uniquely exists.
  There are $\alpha_1,\dots,\alpha_n\in\mathbb{R}$ such that
  \begin{eqnarray}\label{theorem-empirical-representer-theorem-2}
    f_{A,D_n,\lambda}(\cdot)\;=\;
    \sum_{i=1}^n\alpha_i\!\cdot\!\big(A\Phi(\cdot)\big)(z_i)\;.
  \end{eqnarray}
  The matrix $M\in\mathbb{R}^{n\times n}$ defined by
  \begin{eqnarray}\label{theorem-empirical-representer-theorem-3}
    M_{i,j}\;=\;\bigg(A\Big(\big(A\Phi(\cdot)\big)(z_i)\Big)\bigg)(z_j)
  \end{eqnarray}
  is symmetric and positive semi-definite. For every
  $\alpha=(\alpha_1,\dots,\alpha_n)^{\mathsf{T}}
   \in\mathbb{R}^n$,
  \begin{eqnarray}\label{theorem-empirical-representer-theorem-1}
    \mathcal{R}_{A,D,\lambda}
      \bigg(\sum_{i=1}^n\alpha_i\!\cdot\!\big(A\Phi(\cdot)\big)(z_i)
      \bigg)
    \;=\;\frac{1}{n}\sum_{i=1}^n 
            L\big(z_i,y_i,\alpha^{\mathsf{T}}Me_i\big)
         +\lambda\alpha^{\mathsf{T}}M\alpha\;
  \end{eqnarray}
  where $e_i$ denotes the $i$-th vector in the standard basis
  of $\mathbb{R}^n$.
\end{theorem}
Theorem \ref{theorem-empirical-representer-theorem} is of great practical
importance because it says that estimates can be calculated essentially
by finding a minimizer of
$$\alpha\;\mapsto\;\frac{1}{n}\sum_{i=1}^n 
            L\big(z_i,y_i,\alpha^{\mathsf{T}}Me_i\big)
         +\lambda\alpha^{\mathsf{T}}M\alpha
$$
in $\mathbb{R}^n$ where $M$ is a symmetric and positive semi-definite
matrix. This is extremely comfortable because calculating ordinary
regularized kernel methods leads to an optimization problem with 
exactly the same structure. Accordingly, developing new software 
for inverse problems
is unnecessary because, after calculating the matrix $M$, one
can use standard software for regularized kernel methods such as
the R-package ``kernlab'' \citep{kernlab}.
In order to calculate $M$, one only needs to write an R-function
which takes a function $f$ as an argument and returns the function $Af$. 

\medskip

Almost all articles on inverse regression problems assume the 
homoscedastic regression model 
\begin{eqnarray}\label{homoscedastic-regression-model}
  Y\;=\;\big(Af_{A,P}\big)(Z)\,+\,\varepsilon\,.
\end{eqnarray}
Instead of only considering such a specific model,
we use a suitable loss function $L$ and consider the risk
$$\mathcal{R}_{A,P}(f)\;:=\;\int L\big(z,y,(Af)(z)\big)\,P\big(d(z,y)\big)
$$ 
for functions $f:\mathcal{X}\rightarrow\mathbb{R}$ (in the domain of $A$).
Then, the goal is to estimate a minimizer $f_{A,P}$ of this risk.
In this way, it is, e.g., possible to investigate heteroscedastic inverse regression
problems such as 
\begin{eqnarray}\label{heteroscedastic-regression-model}
  Y\;=\;\big(Af_{A,P}\big)(Z)\,+\,s(Z)\varepsilon,
\end{eqnarray}
where $s$ is an unknown scale function or, even more general,
$$Y\;=\;\big(Af_{A,P}\big)(Z)\,+\,\varepsilon_Z,
$$
where $(\omega,z)\mapsto \varepsilon_z(\omega)$ is a Markov kernel. 
In most parts of the article, we do not make any specific assumption on
the regression model but consider minimizing risks.
As a theoretical tool, we also need the regularized risk
$$\mathcal{R}_{A,P,\lambda}(f)\,=\,\mathcal{R}_{A,P}(f)+\lambda\|f\|_H^2
  \,=\,\int\!\! L\big(z,y,(Af)(z)\big)\,P\big(d(z,y)\big)+\lambda\|f\|_H^2
$$ 
for $f\in H$ and $\lambda\in(0,\infty)$. This regularized risk is the 
theoretical counterpart of the regularized empirical risk $\mathcal{R}_{A,D,\lambda}$.
The following theorem presents a simple condition under which a uinique minimizer
of the regularized risk exists. By choosing the empirical measure for $P$,
the theorem also guarantees existence of the estimates $f_{A,D,\lambda}$.
\begin{theorem}\label{theorem-unique-existence-theoretical-svm}
  Let Assumption \ref{assumption-general-setting} be fulfilled and
  \begin{eqnarray}\label{theorem-unique-existence-theoretical-svm-1}
    \int L(z,y,0)\,P\big(d(z,y)\big)\;<\;\infty\,.
  \end{eqnarray}
  Then, for every $\lambda>0$,
  there is a unique minimizer $f_{A,P,\lambda}$ of
  $f\mapsto\mathcal{R}_{A,P,\lambda}(f)$ in $H$.
\end{theorem}

\section{Consistency and Rate of Convergence}\label{section-consistency}

Define i.i.d.\ random variables
$$(X_i,Y_i)\;\sim\;P,
  \qquad i\in\mathbb{N}\,.
$$
Then, the data set $D_n$ is a realization of the random vector
$$\mathbf{D}_n\;=\;
  \big((X_1,Y_1),\dots,(X_n,Y_n)\big)\,.
$$
The following theorem guarantees consistency of regularized kernel methods
for inverse regression problems under quite weak assumptions.
In particular, it also covers the multivariate case as 
$\mathcal{X}$ may be any compact subset of $\mathcal{R}^d$, it does not
require homoscedasticity or any signal plus noise assumption, and 
we do
not assume injectivity of the operator $A$ or any properties of 
a singular system of $A$.
\begin{theorem}\label{theorem-consistency}
  Let Assumptions \ref{assumption-general-setting} and \ref{assumption-on-L}
  be fulfilled,
  and let $A$ fulfill (\ref{assumption-operator-A-continuous-on-Cb}).
  Assume that 
  \begin{eqnarray}\label{theorem-consistency-1}
    \exists\,f^\ast \in H\text{ s.t. }
    \mathcal{R}_{A,P}(f^\ast)\;=\;\inf_{f\in H}\mathcal{R}_{A,P}(f)
    \;:=\;\mathcal{R}_{A,P}^\ast\,.
  \end{eqnarray}
  Then, there is an $f_{A,P}\in H$ such that
  \begin{eqnarray}\label{theorem-consistency-2}
    \mathcal{R}_{A,P}(f_{A,P})\;=\;\inf_{f\in H}\mathcal{R}_{A,P}(f)
  \end{eqnarray}
  and, for every sequence 
  $(\lambda_n)_{n\in\mathbb{N}}\subset(0,\infty)$ such that
  $\lim_{n\rightarrow\infty}\lambda_n^{1+p/2}\sqrt{n}=\infty$, 
  \begin{eqnarray}\label{theorem-consistency-3}
    \big\|f_{A,\mathbf{D}_n,\lambda_n}-f_{A,P}\big\|_H
    \;\;\xrightarrow[\;n\rightarrow\infty\;]{}\;\;0
    \qquad\text{in probability}\;.
  \end{eqnarray}
\end{theorem}
Note that (\ref{theorem-consistency-3}) also implies
$$\sup_{x\in\mathcal{X}}\big|f_{A,\mathbf{D}_n,\lambda_n}(x)-f_{A,P}(x)\big|
    \;\;\xrightarrow[\;n\rightarrow\infty\;]{}\;\;0
    \qquad\text{in probability}\;.
$$

As already mentioned above, Theorem \ref{theorem-consistency} does not require 
any signal plus noise assumption. Instead, it is only assumed that $H$ contains
a minimizer of the risk. In order to make this assumption more explicit
in case of a signal plus noise assumption,
it is exemplified for special choices of the loss function L and the RKHS $H$.
Consider the heteroscedastic model
\begin{eqnarray}\label{heteroscedastic-model-example-consistency}
  Y\;=\;Af_0(Z)\,+\,s(Z)\varepsilon
\end{eqnarray}
where $s$ is an unknown scale function and
$\mathbb{E}\varepsilon=0$ in case of the least-squares loss
or $\text{median}(\varepsilon)=0$ in case of the absolute deviation loss.
(In the latter case, it is additionally assumed that 0 is the unique median of
$\varepsilon$;  
error distributions which violate this assumption are extremely
unusual.)
Then, Assumption (\ref{theorem-consistency-1}) is fulfilled if
$f_0\in H$. As we have not assumed injectivity of $A$ so far, model
(\ref{heteroscedastic-model-example-consistency}) is not necessarily unique.
It is possible that $f_0\not= f_{A,P}$ but it follows from
(\ref{theorem-consistency-2}) that  
$Af_{A,P}=Af_0$ $P_{\mathcal{Z}}$\,-\,a.s.\ so that model
(\ref{heteroscedastic-model-example-consistency}) can be rewritten as 
\begin{eqnarray}\label{heteroscedastic-model-example-consistency-variant}
  Y\;=\;Af_{A,P}(Z)\,+\,s(Z)\varepsilon\,;
\end{eqnarray}
Obviously, distinguishing between models
(\ref{heteroscedastic-model-example-consistency}) and
(\ref{heteroscedastic-model-example-consistency-variant})
is impossible.
In order to prove rates of convergence for $f_{A,\mathbf{D}_n,\lambda_n}-f_{A,P}$
below, we will also assume that $A$ is injective. Under this standard assumption,
the model is unique and $f_0=f_{A,P}$. \\
In a parametric setting, it is typically assumed that $f_0$ 
is linear or a polynomial. This 
assumption easily implies
(\ref{theorem-consistency-1}) if $k$ is the linear kernel or a suitable 
polynomial kernel. In a nonparametric setting, the most common kernel
is the Gaussian RBF kernel. However, assuming that a function $f_0$
lies in the corresponding RKHS $H$ is a rather strong and inaccessible assumption
which can hardly be made explicit for a practitioner -- even though this RKHS
is dense in $\mathcal{C}(\mathcal{X})$. Therefore, it seems advisable to 
use slightly different kernels in the nonparametric setting, namely
Wendland kernels. These are radial kernels of the form
$$k_{d,m}(x,x^\prime)=\phi_{d,m}\big(\|x-x^\prime\|\big)
  \quad\text{with}\quad
  \phi_{d,m}(r)=
  \left\{
     \begin{array}{ll}
		p_{d,m}(r), \;\; & 0\leq r\leq 1 \\
		0, & r>1
     \end{array}
  \right.
$$
where $p_{d,m}$ is a certain polynomial of degree $\lfloor d/2 \rfloor+3m+1$.
The polynomial is of minimal degree such that $k_{d,m}$ is 
$m$-times continuously differentiable; see \cite[Theorem 9.12 and 9.13]{wendland2005}.
Though the shape of these kernels is very similar to that of the Gaussian RBF kernel,
Wendland kernels have two advantages: First, they are compactly supported and therefore
lead to sparse kernel matrices. Second, there is a simple condition on 
$f_{0}$ which guaranties that $f_{0}$ is contained in the corresponding
RKHS $H_{d,m}$, i.e., that (\ref{theorem-consistency-1}) is fulfilled.  
If the dimension $d$ is odd, choose $m=(d+1)/2$ and $\gamma=d+1$. If
$d$ is even, choose $m=d/2+1$ and $\gamma=d+2$. Then, $f_{0}$
is in $H_{d,m}$ if it is the restriction of a $\gamma$-times continuously differentiable
function on $\mathbb{R}^d$. This is a consequence of the fact that
the RKHS of $k_{d,m}$ is the Sobolev space $H^{d/2+m+1/2}(\mathbb{R}^d)$,
that $\gamma\geq d/2+m+1/2$,
and that $\mathcal{X}$ is bounded; see \cite[Theorem 10.35 and \S\,10.7]{wendland2005}.
For the convenience of the reader, Table 
\ref{table-wendland-polynomials} contains the relevant Wendland 
polynomials $p_{d,m}$ for dimensions $d\leq 5$; these are calculated from 
\cite[Cor.\ 9.15]{wendland2005}. Polynomials $p_{d,m}$ for higher dimensions
can recursively be obtained from \cite[Theorem 9.12]{wendland2005}. 
\begin{table}\renewcommand{\arraystretch}{1.3}
\begin{tabular}{ll}
	Dimension & Wendland polynomial \\ \hline
	$d=1$ & $p_{1,1}(r)\;=\;(1-r)^3(3r+1)$ \\
	$d=2$ & $p_{2,2}(r)\;=\;(1-r)^6(35r^2+18r+3)$ \\
	$d=3$ & $p_{3,2}(r)\;=\;(1-r)^6(35r^2+18r+3)$ \\
	$d=4$ & $p_{4,3}(r)\;=\;(1-r)^9(693r^3+477r^2+135r+15)$ \\
	$d=5$ & $p_{5,3}(r)\;=\;(1-r)^9(693r^3+477r^2+135r+15)$ \\
\end{tabular}
\caption{Suitable Wendland polynomials $p_{d,m}$ 
         for different dimensions $d$; that is, $m=(d+1)/2$ if $d$ is odd,
         $m=d/2+1$ if $d$ is even.
}
\label{table-wendland-polynomials}
\end{table}

\bigskip

For the risk of the estimator, we obtain a rate of convergence 
-- even under the quite weak assumptions of Theorem \ref{theorem-consistency}.
For this rate, neither an assumption 
on the singular system of $A$ nor a signal plus noise assumption such as
(\ref{homoscedastic-regression-model}) or (\ref{heteroscedastic-regression-model})
is needed. 
\begin{theorem}\label{theorem-rate-of-convergence-risks}
  Let Assumptions \ref{assumption-general-setting} and \ref{assumption-on-L}
  be fulfilled, let
  $A$ fulfill (\ref{assumption-operator-A-continuous-on-Cb}),
  and assume (\ref{theorem-consistency-1}).
  Let $(\lambda_n)_{n\in\mathbb{N}}\subset(0,\infty)$ and 
  $(a_n)_{n\in\mathbb{N}}\subset(0,\infty)$ be sequences
  such that $\lim_{n\rightarrow\infty}\lambda_n=0$,
  $\lim_{n\rightarrow\infty}a_n=\infty$, 
  \begin{eqnarray}\label{theorem-rate-of-convergence-risks-1}
    \lim_{n\rightarrow\infty}
    \frac{a_n}{\lambda_n^{1+p/2}\sqrt{n}\,}
    \;=\;0,
    \qquad\text{and}\qquad
     \lim_{n\rightarrow\infty}\lambda_n a_n\;=\;0.
  \end{eqnarray}
  Then,
  \begin{eqnarray}\label{theorem-rate-of-convergence-risks-2}
    a_n\Big(\mathcal{R}_{A,P}\big(f_{A,\mathbf{D}_n,\lambda_n}\big)
            -\mathcal{R}_{A,P}^\ast
       \Big)
    \;\;\xrightarrow[\;n\rightarrow\infty\;]{}\;\;0
    \qquad\text{in probability}.\;.
  \end{eqnarray}
  In particular, if $\lambda_n=\gamma n^{-1/(4+p)}\;\,\forall\,n\in\mathbb{N}$
  for some constant $\gamma\in(0,\infty)$, then every sequence
  $(a_n)_{n\in\mathbb{N}}\subset(0,\infty)$ such that
  $$\lim_{n\rightarrow\infty}a_n n^{-1/(4+p)}\;=\;0
  $$  
  fulfills condition (\ref{theorem-rate-of-convergence-risks-1}). 
\end{theorem}

\bigskip

In the rest of this section, we are concerned with rates of convergence for
$f_{A,\mathbf{D}_n,\lambda_n}-f_{A,P}$. While
Theorems \ref{theorem-empirical-representer-theorem},
\ref{theorem-unique-existence-theoretical-svm},
\ref{theorem-consistency}, and
\ref{theorem-rate-of-convergence-risks} are valid
for all of the commonly used loss functions and do not require
any involved assumptions on the distribution $P$ and the operator $A$,
obtaining rates of convergences for $f_{A,\mathbf{D}_n,\lambda_n}-f_{A,P}$
is certainly only possible under
much more restrictive and involved assumptions. Therefore, we need some
preparations, before we are able to state such rates of convergence in 
Theorem \ref{theorem-rate-of-convergence} below. 

\medskip

\textit{Loss function $L$.}

Nearly all
articles on inverse (regression) problems which 
(at least implicitly) employ a loss function are restricted to the least-squares
loss. (A notably exception is \cite{krebs2011} which uses the
$\varepsilon$-insensitive loss common in machine learning.)
While we did not have to specify a special loss function so far,
we fix a specific loss function now, namely the absolute deviation loss.
That is, our loss function is
\begin{eqnarray}\label{absolute-deviation-loss}
  L\;:\;\;\mathcal{Z}\times\mathcal{Y}\times\mathbb{R}\;\rightarrow\;[0,\infty),
  \qquad (z,y,t)\;\mapsto\;|y-t|
\end{eqnarray} 
in the following. On the one hand, this choice of the loss function
is motivated by the fact that the absolute deviation loss 
typically leads to better robustness properties than the least-squares loss.
In particular, it is shown in \cite{HableChristmann2011}, 
\cite{christmannsteinwart2007}, and 
\cite{christmann2009} that ordinary regularized kernel 
methods based on the absolute deviation loss enjoy a qualitative 
robustness property and have
a bounded influence function and a bounded maxbias. On the other hand,
the absolute deviation loss together with
suitable assumptions on the model guarantee a bound of the form
\begin{eqnarray*}\label{form-of-a-bound-functions-and-risks}
  \big\|g-g^\ast\big\|_{L_q(P_{\mathcal{Z}})}\;\leq\;
  \Big(
    \mathcal{R}_{\text{Id},P}\big(g\big)-
    \mathcal{R}_{\text{Id},P}^\ast
  \Big)^r
\end{eqnarray*}  
where 
$\mathcal{R}_{\text{Id},P}(g^\ast)=\int\! L(z,y,g^\ast(z))P(d(z,y))
 =\inf_g \int\! L(z,y,g(z))P(d(z,y))
$
and $q,r\in(0,\infty)$. However, by adapting the model assumptions, 
such bounds can also
be obtained for other loss functions; see \cite[\S\,3.9]{steinwart2008}.

\medskip

\textit{Inverse regression model.}

In the following, we assume the heteroscedastic regression model
\begin{eqnarray}\label{heteroscedastic-regression-model-nochmal}
  Y\;=\;\big(Af_{A,P}\big)(Z)\,+\,s(Z)\varepsilon,
\end{eqnarray}
where $s$ is an unknown scale function such that
\begin{eqnarray}\label{heteroscedastic-regression-model-assumption-scale-function}
  \text{there are constants }
  \underline{c}_s,\overline{c}_s\in(0,\infty)\;\text{ with }
  \;\underline{c}_s\leq s\leq \overline{c}_s,
\end{eqnarray}
the random error $\varepsilon$ is independent
from $Z$, has
\begin{eqnarray}\label{heteroscedastic-regression-model-assumption-error-median}
  \text{median}(\varepsilon)\;=\;0,
\end{eqnarray}
and
\begin{eqnarray}\label{heteroscedastic-regression-model-assumption-error-density-1}
  & &
  \text{ the distribution of }\varepsilon\text{ has a Lebesgue-density }h
  \text{ such that}\qquad\qquad\qquad \\
  & & \label{heteroscedastic-regression-model-assumption-error-density-2}
  \quad\exists\,a_h,c_h\,\in\,(0,\infty)\;\text{ s.t.\ }\;\forall\,y\in(-a_h,a_h)\,:\;
  h(y)\geq c_h \,.
\end{eqnarray}
If the distribution of the error $\varepsilon$ has a Lebesgue-density, then
Assumption (\ref{heteroscedastic-regression-model-assumption-error-density-2})
is very natural and, e.g., fulfilled for all
unimodal error distributions. If $h$ is continuous, it is sufficient
that $h(0)\not=0$.

\medskip

\textit{Operator $A$.}

In order to obtain rates of convergences for 
$f_{A,\mathbf{D}_n,\lambda_n}-f_{A,P}$, we will also need the standard assumption that
\begin{eqnarray}\label{assumption-operator-injective-1}
  A:\;H\;\rightarrow\;\mathcal{C}_b(\mathcal{Z})
  \quad\text{is injective}.
\end{eqnarray}
Let $\mathcal{Z}_0$ be the support of $P_{\mathcal{Z}}$ and let
$P_{\mathcal{Z}_0}$ denote the restriction of $P_{\mathcal{Z}}$
on $\mathcal{Z}_0$. Then, the natural embedding 
$\iota_0:\mathcal{C}_b(\mathcal{Z})\rightarrow L_2\big(P_{\mathcal{Z}_0}\big)$
defines a continuous linear operator $A_0:=\iota_0\circ A$ and it is easy to see
that (\ref{assumption-operator-injective-1}) implies that
\begin{eqnarray}\label{assumption-operator-injective-2}
  A_0:\;H\;\rightarrow\;L_2\big(P_{\mathcal{Z}_0}\big)
  \quad\text{is injective}.
\end{eqnarray}
Furthermore, it follows from Assumption 
(\ref{assumption-operator-A-continuous-on-Cb}) and Prop.\
\ref{prop-compactness-A} that
$A_0$ is a compact operator. In this case,
$A_0$ has a singular system $(\sigma_j;v_j,u_j)_{j\in\mathbb{N}}$;
see, e.g., \cite[\S\,2.2]{Engl:Hanke:1996}.
That is,
$\sigma_j^2$, $j\in\mathbb{N}$, are the non-zero eigenvalues of the self-adjoint
operator $A_0^\ast A_0$ such that
$\sigma_1\geq \sigma_2\geq \dots >0$. The set $\{v_j\,|\,j\in\mathbb{N}\}$ 
is a corresponding complete orthonormal system of $H$; completeness follows from
injectivity of $A_0$ and (\ref{singular-value-decomposition-2})
below. Finally,
the elements
$$u_j\;:=\;\frac{A_0v_j}{\;\|A_0v_j\|_{L_2(P_{\mathcal{Z}_0})}}\,,
  \qquad j\in\mathbb{N},
$$
form a complete orthonormal system of the closure of
$\{A_0f\,|\,f\in H\}$ in $L_2\big(P_{\mathcal{Z}_0}\big)$.
We have
\begin{eqnarray}
 &&\label{singular-value-decomposition-1}
  A_0v_j\;=\;\sigma_j u_j\,,\qquad
  A_0^\ast u_j\;=\;\sigma_j v_j
  \qquad\forall\,j\in\mathbb{N} \\
 &&\label{singular-value-decomposition-2}
  A_0 f\;=\;\sum_{j=1}^\infty \sigma_j\langle f,v_j\rangle_{H}u_j
  \qquad\forall\,f\in H \\
 && \label{A-adjoint-expansion}
  A_0^\ast g\;=\;
    \sum_{j=1}^\infty \sigma_j\langle g,u_j\rangle_{L_2(P_{\mathcal{Z}_0})}v_j
  \qquad\forall\,g\in L_2(P_{\mathcal{Z}_0})\,. 
\end{eqnarray}

Now, we can state the theorem on rates of convergence for
$f_{A,\mathbf{D}_n,\lambda_n}-f_{A,P}$. 
\begin{theorem}\label{theorem-rate-of-convergence}
  Let Assumption \ref{assumption-general-setting} be fulfilled
  and assume that the marginal distribution $P_{\mathcal{Y}}$ has a finite first moment,
  i.e., $\mathbb{E}|Y|=\int |y|\,P(d(z,y))<\infty$.
  Let $L$ be the absolute deviation loss 
  (\ref{absolute-deviation-loss}) and assume the heteroscedastic
  model given by 
  (\ref{heteroscedastic-regression-model-nochmal})--%
  (\ref{heteroscedastic-regression-model-assumption-error-density-2}).
  Let $A$ fulfill (\ref{assumption-operator-A-continuous-on-Cb}),
  and (\ref{assumption-operator-injective-1}). \\
  Let $(\lambda_n)_{n\in\mathbb{N}}\subset(0,\infty)$ and 
  $(a_n)_{n\in\mathbb{N}}\subset(0,\infty)$ be sequences
  with $\lim_{n\rightarrow\infty}\lambda_n=0$,
  $\lim_{n\rightarrow\infty}a_n=\infty$, and 
  \begin{eqnarray}\label{theorem-rate-of-convergence-1}
    \lim_{n\rightarrow\infty}
    \frac{\sqrt{a_n}}{\lambda_n\sqrt{n}\,}
    \;=\;0.
  \end{eqnarray}
  Then, the following assertions are valid:
  \begin{enumerate}
   \item[(a)] If
     \begin{eqnarray}\label{theorem-rate-of-convergence-2}
        \sum_{j=1}^\infty 
         \frac{\big|\big\langle f_{A,P},v_j\big\rangle_{\!H}\big|^2}{\sigma_j^2}
       \;<\;\infty
     \end{eqnarray}
     and $\,\lim_{n\rightarrow\infty}a_n\lambda_n^{\frac{1}{2}}=0\,$,
     then  
     \begin{eqnarray}\label{theorem-rate-of-convergence-4}
       a_n
          \big\|f_{A,\mathbf{D}_n,\lambda_n}
                -f_{A,P}
          \big\|_H^2
       \;\;\xrightarrow[\;n\rightarrow\infty\;]{}\;\;0
       \qquad\text{in probability}.\quad
     \end{eqnarray}  
   \item[(b)] Fix any $x\in\mathcal{X}$. If  
        \begin{eqnarray}\label{theorem-rate-of-convergence-7}
           \sum_{j=1}^\infty 
            \frac{\big(v_j(x)\big)^2}{\sigma_j^2}
            \;<\;\infty
        \end{eqnarray}
        and $\,\lim_{n\rightarrow\infty}a_n\lambda_n=0\,$,
        then  
        \begin{eqnarray}\label{theorem-rate-of-convergence-6}
          a_n\big(f_{A,\mathbf{D}_n,\lambda_n}(x)
                   -f_{A,P}(x)
             \big)^2
          \;\;\xrightarrow[\;n\rightarrow\infty\;]{}\;\;0
          \qquad\text{in probability}.\quad
        \end{eqnarray}
 \end{enumerate}  
\end{theorem}
  For example, in part (a), let 
  $\lambda_n=\gamma n^{-2/5}\;\,\forall\,n\in\mathbb{N}$
  for some constant $\gamma\in(0,\infty)$; then all conditions on
  $(a_n)_{n\in\mathbb{N}}\subset(0,\infty)$ are fulfilled
  if $\lim_{n\rightarrow\infty}a_n=\infty$ and
  $$\lim_{n\rightarrow\infty}a_n n^{-1/5}\;=\;0\,.
  $$
  In part (b), let 
  $\lambda_n=\gamma n^{-1/3}\;\,\forall\,n\in\mathbb{N}$
  for some constant $\gamma\in(0,\infty)$; then all conditions on
  $(a_n)_{n\in\mathbb{N}}\subset(0,\infty)$ are fulfilled
  if $\lim_{n\rightarrow\infty}a_n=\infty$ and
  $$\lim_{n\rightarrow\infty}a_n n^{-1/3}\;=\;0\,.
  $$

Assumptions such as (\ref{theorem-rate-of-convergence-2}) in part (a) 
are often called smoothness
assumptions on $f_{A,P}$ and are common in order to obtain rates of convergence.
Assumption (\ref{theorem-rate-of-convergence-7}) in part (b) differs as it does not
involve $f_{A,P}$ but is a condition on a fixed $x$. The result 
of part (b) can be interpreted in the following way: if the 
inverse regression problem is only moderately ill-posed in some area, then
any $f_{A,P}$ can be estimated with a certain rate of convergence in that
area. Of course, the convergence in (\ref{theorem-rate-of-convergence-4}) and
(\ref{theorem-rate-of-convergence-6}) could also be reformulated without
the exponent 2. However, this formulation makes it possible to easily compare
the results with other rates of convergence which most often apply to
$$\mathbb{E}\big\|f_{A,\mathbf{D}_n,\lambda_n}
                  -f_{A,P}
            \big\|_\star^2
$$ 
where $\|\cdot\|_\star$ denotes the respectively considered norm.

\section{Simulation}\label{section-simulation}

In order to illustrate the use of regularized kernel methods for 
inverse regression problems, this section contains simulations in a typical
example of an ill-posed problem, namely backward heat conduction. In a statistical
setting, this example has also been considered, e.g., in 
\cite{Mair:Ruymgaart:1996} and \cite{Bissantz:Holzmann:2013}.
According to, e.g.\ \cite[Example 15.3]{kress:1999}, we are faced with the 
heat equation
$$\frac{\partial u}{\partial t}\;=\;\frac{\partial^2 u}{\partial^2 x}
$$
where
$$u\,:\;[0,1]\times [0,T]\;\rightarrow\;\mathbb{R}\,,\qquad
  (x,t)\;\mapsto\;u(x,t)
$$
denotes the temperature at any spatial point $x\in[0,1]$ and time
$t\in[0,T]$. The boundary conditions are $u(0,t)=u(1,t)=0$ for every $t\in[0,T]$
and the goal is to recover the initial conditions
$$f(x)\;:=\;u(x,0),\qquad x\in[0,1],
$$
at time $t=0$ from (noisy) observations at time $t=T$. Let
$A$ denote the operator which maps the initial state $f$ to the final
temperature curve $g=u(\cdot,T)$ at time $T$. Then,
\begin{equation}\label{simulation-def-operator}
  \big(Af\big)(z)\;=\;u(z,T)\;=\;
  \sum_{j=1}^\infty \exp(-j^2\pi^2 T)
    \!\int_{[0,1]} f(x)v_j(x)\lambda(dx)\,v_j(z)\;\;
\end{equation} 
with $v_j(z)=\sqrt{2}\sin(j\pi z)$ for every $z\in[0,1]$.
That is $\mathcal{X}=\mathcal{Z}=[0,1]$ and $(v_j)_{j\in\mathbb{N}}$
is a complete orthonormal system of $L_2([0,1])$. As a mapping from
$L_2([0,1])$ to $L_2([0,1])$, the operator $A$ 
is self-adjoint with
eigenfunctions $v_j$ and eigenvalues $\sigma_j=\exp(-j^2\pi^2 T)$.
We use this standard example so that the method can be compared
to the spectral cut-off estimator suggested by \cite{Mair:Ruymgaart:1996}.

\medskip

\begin{figure}
\begin{center}
  \includegraphics[width=0.5\textwidth]{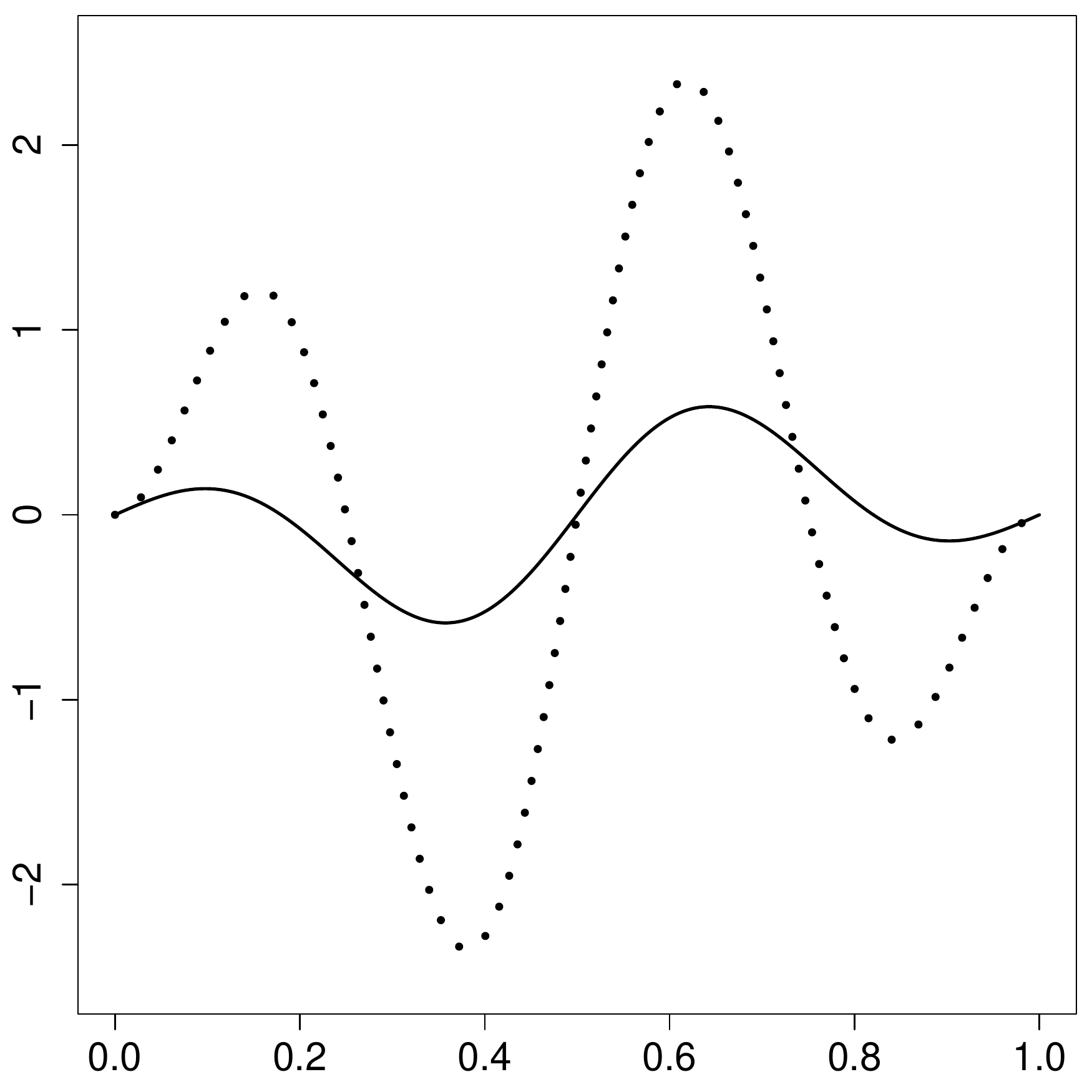}
  \caption{The regression function $A(f_0)$ (solid line) and 
           the function $f_0$ (dotted line) which has to be recovered.}
  \label{fig-modelfunction}
\end{center}
\end{figure}

\textit{The model.} 
We simulate data
\begin{equation}\label{simulation-model-1}
  y_i\;=\;\big(Af_0\big)(z_i)+s_\delta(z_i)\varepsilon_i\,,\qquad i\in\{1,\dots,n\},
\end{equation} 
with
\begin{equation}\label{simulation-model-2}
  f_0(x)\,=\,-10x(x-1)\sin(4\pi x)\qquad\text{and}\qquad
  s_\delta(z)\,=\,\delta z(z-1)\qquad
\end{equation}
for $T=0.01$ and
different values of the scale factor $\delta$.
The regression function $A(f_0)$ and the function $f_0$ we want to recover
are shown in Figure~\ref{fig-modelfunction}.
We use fixed equidistant design points $z_i=(i-1)/n$ on $[0,1]$ because
most theoretical results on spectral cut-off estimators are for such design points.
This clearly favors the spectral cut-off estimator also because this
estimator is heavily based on the knowledge that the $z_i$ are uniform and
the regularized kernel method does not need (and use) such information.
The errors $\varepsilon_i$ are (independently) sampled from the standard normal
distribution. We consider the three scale factors $\delta\in\{0.25,0.5,1\}$
which result in errors of similar sizes as 
in \cite[\S\,3.2]{Bissantz:Holzmann:2013}. The sample sizes are 
$n\in\{100,250,500,1000\}$.

\medskip

\textit{The estimators.}
As estimators, we use a regularized kernel method (RKM) and a spectral
cut-off estimator (SCE). In case of the RKM, we choose 
the absolute deviation loss function and the
rescaled Wendland
kernel 
$$k_{1,1}^{(0.3)}(x,x^\prime)\;=\;\phi_{1,1}\big(|x-x^\prime|/0.3\big)
$$
with $\phi_{1,1}(r)=(1-r)_+^3(3r+1)$ for every $r\in[0,\infty)$;
see \cite[Table 9.1]{wendland2005}.
The scaling 0.3 is approximately equal to the median of the $n^2$ values
$|x_i-x_j|$, $i.j\in\{1,\dots,n\}$.
This heuristic is the analogue of a heuristic
used in \cite{caputo2002} and \cite[p.\ 9]{kernlab}
in order to choose the scaling factor in case of the Gaussian RBF kernel.
The regularization parameter is equal to $\lambda=\frac{1}{2}an^{-0.45}$ where
$a$ is selected via a 5-fold cross validation among the values
$$10^{-4},\; 5\cdot 10^{-4},\; 10^{-3},\; 5\cdot 10^{-3},\; 10^{-2},\; 
    5\cdot 10^{-2},\; 10^{-1}
$$
in each run of the simulation.
According to \cite[\S\,7.2]{Mair:Ruymgaart:1996},
the spectral cut-off estimator (SCE) is given by
$$\hat{f}_{n,J}(x)\;=\;
  \sum_{j=1}^J \frac{\hat{b}_{j,n}}{\sigma_j}v_j(x)
  \qquad\text{with}\quad \hat{b}_{j,n}=\frac{1}{n}\sum_{i=1}^n v_j(z_i)y_i
$$
where the number $J\in\mathbb{N}$ of basis functions is a regularization parameter 
which is selected via a 5-fold cross validation among the values
$$2,\;3,\;4,\;5,\,\dots\,,\;10
$$
in each run of the simulation.

\medskip

\begin{figure}
\begin{center}
  \includegraphics[width=0.9\textwidth]{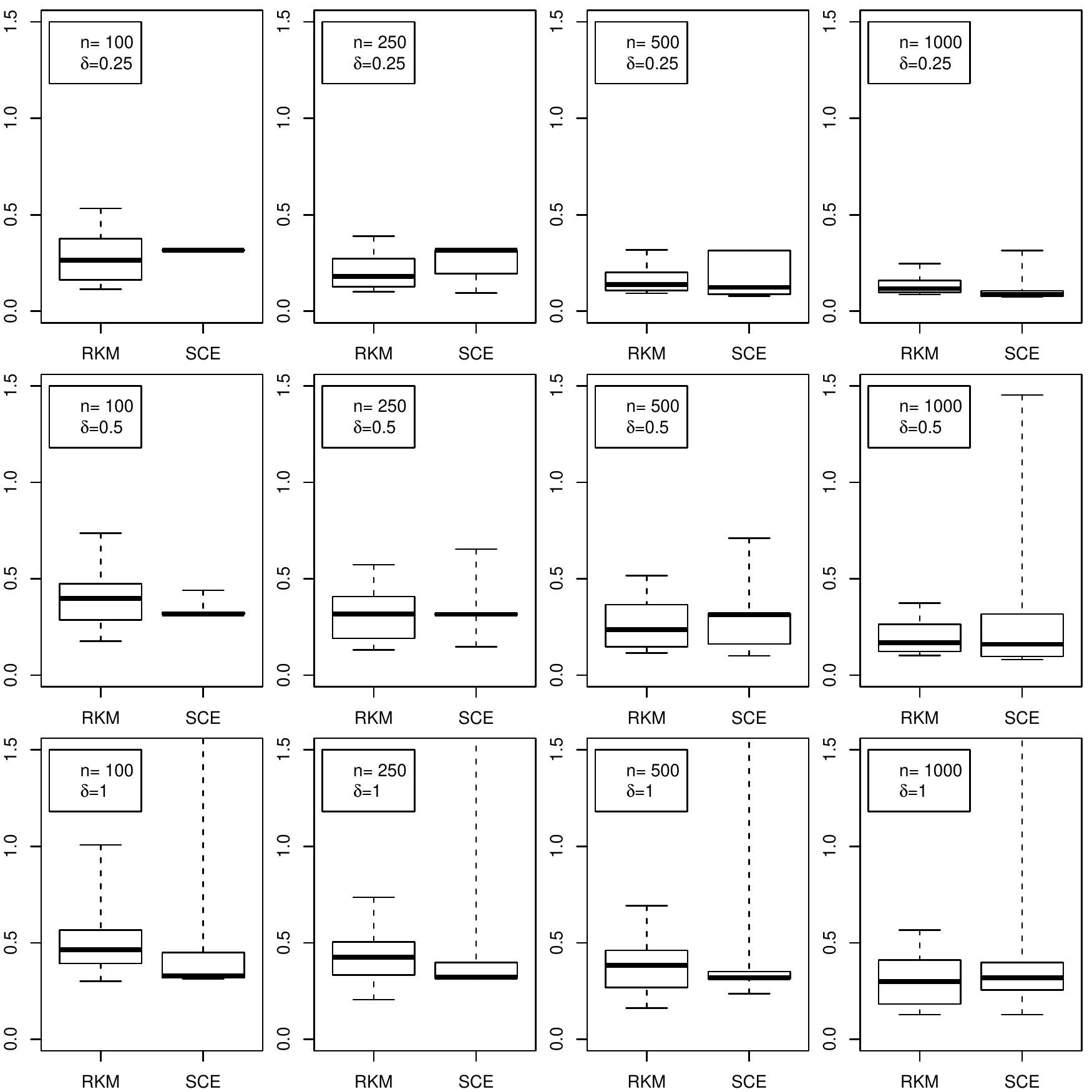}
  \caption{The boxplots of the results $b_1,\dots,b_{1000}$ for both estimators
           in each situation; the ends of the whiskers represent the 10 and 
           the 90 percent quantiles respectively.}
  \label{fig-boxplots}
\end{center}
\end{figure}

\textit{Performance results.}
The simulation consists of $1000$ runs. In each run $r\in\{1,\dots,1000\}$, 
the quality of the 
estimate $\hat{f}_n^{(r)}$ is measured by
$$b_r\;=\;\int_{[0,1]} \big|f_0(x)-\hat{f}_n^{(r)}(x)\big|\,\lambda(dx)\;.
$$
The medians and the boxplots 
(the ends of the whiskers represent the 10 and the 90 percent quantiles respectively)
of the values $b_1,\dots,b_{1000}$ for both estimators
are shown in Table \ref{table-medians} and Figure \ref{fig-boxplots}, 
respectively, in each situation.  
\begin{table} 
\begin{center}
\begin{tabular}{l||cc|cc|cc|cc}
  & \multicolumn{2}{c|}{$n=100$} & \multicolumn{2}{c|}{$n=250$}
  & \multicolumn{2}{c|}{$n=500$} & \multicolumn{2}{c}{$n=1000$} \\
  $\;\;\delta$ & \text{RKM} & \text{SCE} & \text{RKM} & \text{SCE} 
  & \text{RKM} & \text{SCE} & \text{RKM} & \text{SCE} \\ \hline
  0.25 & 0.26 & 0.32 & 0.18 & 0.32 & 0.14 & 0.12 & 0.12 & 0.09 \\
  0.5  & 0.40 & 0.32 & 0.32 & 0.32 & 0.24 & 0.32 & 0.17 & 0.16 \\
  1    & 0.46 & 0.33 & 0.42 & 0.32 & 0.38 & 0.32 & 0.30 & 0.32 \\
\end{tabular}
  \caption{The medians of the results $b_1,\dots,b_{1000}$ for both estimators
           in each situation.}
  \label{table-medians}
\end{center}
\end{table}
In most cases, the performance of both estimators is similar.
However, a closer look on the boxplots reveals that, in case of the SCE, the
25, 50, and 75 percent quantiles are very close to each other but the 90 percent
quantile is far off for some values of $n$ and $\delta$. The reason for this is
that the estimate is very stable for small values of $J$ such as $J=4$ or $J=5$
and the 5-fold cross validation most often chooses these values in some situation.
However, the SCE turns out to be very sensitive to the choice of $J$ in our 
simulations.
If larger values such as $J\in\{6,7,8\}$ are selected, than there is a considerable 
danger that the SCE breaks down. 
Therefore, selecting the right value $J$ is crucial and automatic 
data-driven methods such as a $k$-fold cross validation might not be sufficient.
Table \ref{table-90percent-quantiles} shows the 90 percent quantiles of
the values $b_1,\dots,b_{1000}$ for both estimators in each situation.
From these quantiles, it can be seen that
the SCE frequently breaks down for larger values of $n$ and $\delta$.
Therefore, Table \ref{table-medians} does not show the mean but the median of 
the values $b_1,\dots,b_{1000}$ because the mean is corrupted by large 
outliers in case of the SCE. As this does not happen in case of the RKM,
using the median favors the SCE.
\begin{table} 
\begin{center}
\begin{tabular}{l||cc|cc|cc|cc}
  & \multicolumn{2}{c|}{$n=100$} & \multicolumn{2}{c|}{$n=250$}
  & \multicolumn{2}{c|}{$n=500$} & \multicolumn{2}{c}{$n=1000$} \\
  $\;\;\delta$ & \text{RKM} & \text{SCE} & \text{RKM} & \text{SCE} 
  & \text{RKM} & \text{SCE} & \text{RKM} & \text{SCE} \\ \hline
  0.25 & 0.53 & 0.32 & 0.39 & 0.32 & 0.32 & 0.32 & 0.25 & 0.31 \\
  0.5  & 0.74 & 0.44 & 0.57 & 0.65 & 0.52 & 0.71 & 0.37 & 1.45 \\
  1    & 1.01 & 5.33 & 0.74 & 4.89 & 0.69 & 2.68 & 0.57 & 5.59 \\
\end{tabular}
  \caption{The 90 percent quantiles of the results 
           $b_1,\dots,b_{1000}$ for both estimators
           in each situation.}
  \label{table-90percent-quantiles}
\end{center}
\end{table}

\medskip

\textit{Details on computations.}
The computation of the spectral cut-off estimator 
is simple and extremely fast in this case because the spectral decomposition
is already known here. 
As mentioned
below Theorem \ref{theorem-empirical-representer-theorem}, the computation of
regularized kernel methods for inverse problems can be done by using standard software
for computing ordinary regularized kernel methods.
The only additional thing one has to do is to calculate the 
pseudo kernel matrix $M$ as defined in 
(\ref{theorem-empirical-representer-theorem-3}). Here, we have
$$M_{i,j}\;=\;\sum_{q=1}^\infty\sum_{s=1}^\infty\exp\!\big(\!-\!(q^2+s^2)\pi^2 T\big)
				\mathcal{I}_{q,s}\!\cdot\! v_q(z_j)v_s(z_i)
$$ 
for
$$\mathcal{I}_{q,s}\;=\;
				\int_{[0,1]^2}k_{1,1}^{(0.3)}(x_1,x_2)v_q(x_1)v_s(x_2)\,
                \lambda^2\big(d(x_1,x_2)\big)\,.
$$ 
As the exponential coefficients decrease extremely fast, the infinite
double series can be approximated very well by only calculating a
few terms, e.g., up to $q,s=30$ is more than enough. The double integrals
$\mathcal{I}_{q,s}$
are approximated by a Monte-Carlo simulation in our simulated
example. Then, the coefficients $\alpha_1,\dots,\alpha_n$ are calculated
by the R-package ``kernlab'' (a standard software for calculating ordinary
regularized kernel methods), in the following way:
\begin{eqnarray*}
  &&\texttt{model $\leftarrow$ kqr(as.kernelMatrix(M), y, tau=0.5, C=cost)} \\
  &&\texttt{alpha $\leftarrow$ alpha(model)}
\end{eqnarray*}
where \texttt{y} denotes the vector of the observed values $y_i$, \texttt{tau=0.5}
corresponds to using the absolute deviation loss function, and \texttt{cost}
is the cost regularization parameter which is equal to $1/(2n\lambda)$.   
It is worth mentioning that it is sufficient to calculate
$M$ once and to reuse this matrix in every step of the $k$-fold cross validation
for tuning \texttt{cost}. Finally, values of the estimate $\hat{f}_n$
can be calculated according to (\ref{theorem-empirical-representer-theorem-2});
here, we have
$$\hat{f}_n(x)\;=\;
    \sum_{q=1}^\infty \exp(-q^2\pi^2 T)
      \mathcal{I}_q(x)
      \sum_{i=1}^n \alpha_i v_q(z_i)                       
$$
for 
$$\mathcal{I}_q(x)\;=\;\int_{[0,1]}k_{1,1}^{(0.3)}(x_1,x)v_q(x_1)\,\lambda(dx_1)\,.
$$ 
Again, the extremely fast decay of the exponential coefficients guarantees
that the series can be approximated very well by only calculating a
few terms.

\section{Conclusions}\label{section-conclusions}

Regularized kernel methods constitute a broad and flexible class of
methods which originate from machine learning and are 
common in nonparametric classification and regression problems today.
In this article, we investigated the use of 
regularized kernel methods for inverse problems in a unifying way.
In addition to consistency results under very weak assumptions,
we also obtained a rate of convergence under a typical smoothness assumption
on the target function. Though such a rate of convergence is 
of a purely theoretical manner and is not interesting on its own for real data analysis,
it can play an important role in developing methods for
statistical inference (such as asymptotic confidence sets)
based on undersmoothing.
However, statistical inference is still at an early stage
even in case of regularized kernel methods for ordinary regression problems
as well as in case of inverse regression problems with any other estimation method.
First steps on statistical inference for regularized kernel methods are done
in
\cite{brabanter2011}, 
\cite{hable2012a}, and 
\cite{Hable:2012b}. In case of inverse regression problems,
\cite{Bissantz:Holzmann:2008} and \cite{Bissantz:Birke:2009} 
are concerned with asymptotic confidence sets for spectral cut-off 
estimators. 
Accordingly, enabling statistical inference for inverse regression problems
with regularized kernel methods is a matter of future and challenging research.
Using regularized kernel methods in real data analysis 
of inverse regression problems seems to be promising as
they have nice computational properties and it is possible to
resort to already existing well developed software implementations
of ordinary regularized kernel methods.

\section{Appendix}

\subsection{Additional Results}\label{section-appendix-addiditional-results}

This subsection contains some additional results on regularized 
kernel methods for inverse problems. On the one hand, these results
are needed in the proofs of the main results; on the other hand,
they are also interesting of its own, in particular, as some of them are the 
counterparts
of some of the main tools for ordinary regularized 
kernel methods (i.e. $A=\text{id}$).

The first proposition shows that, in our setting, compactness
of $A$ (in a rather week sense) comes for free by assuming 
(\ref{assumption-operator-A-continuous-on-Cb}).
\begin{proposition}\label{prop-compactness-A}
  Let Assumption \ref{assumption-general-setting} 
  be fulfilled, and
  assume that $A$ fulfills (\ref{assumption-operator-A-continuous-on-Cb}). Then,
  $A:H\rightarrow\mathcal{C}_b(\mathcal{Z})$ is a compact operator; that is,
  \begin{eqnarray}\label{prop-compactness-A-1}
    (f_n)_{n\in \mathbb{N}_0}\subset H,\;\;
    f_n\,\xrightarrow[]{\;\;\;\text{w}\;\;\;}\,f_0
    \quad\;\Rightarrow\;\quad
    \big\|A(f_n)-A(f_0)\big\|_\infty\xrightarrow[]{\;\;\;\;\;\;}\,0
  \end{eqnarray}
  where $\,\,\xrightarrow[]{\;\;\text{w}\;\;}\,\,$ denotes weak convergence in the 
  Hilbert space.
\end{proposition}

The next theorem is a general representer theorem. In case of
ordinary regularized kernel methods, general representer theorems
are \emph{the} main tool for deriving theoretical results on 
consistency, rates of convergence, asymptotic normality, and robustness.
The proof of the following theorem for 
the case of inverse problems is similar to 
the proof of the general representer theorem in the ordinary case
\cite[Theorem 5.8 and Theorem 5.9][see, e.g.,]{steinwart2008}
even though the assumptions and the result considerably differ. 
\begin{theorem}[General Representer Theorem]
  \label{theorem-general-representer} \hfill \\
  Let Assumptions \ref{assumption-general-setting} and \ref{assumption-on-L}
  be fulfilled, and
  fix any $\lambda>0$. 
  Then, there is
  an $h_{P,\lambda}\in\mathcal{L}_2(P)$ with the following properties:
 \begin{enumerate}[leftmargin=2em]
 \item[(a)] For every $(z,y)\in\mathcal{Z}\times\mathcal{Y}$, 
  \begin{eqnarray}\label{theorem-general-representer-1}
    \big|h_{P,\lambda}(z,y)\big|\!\!\!&\leq&\!\!\!
    b_0^\prime(y)+
    b_1^\prime\!\cdot\!\!\left(\!\|A\|\sqrt{\frac{1}{\lambda}\int b\,dP\,}
                                +1\!
                   \right)^{p},\qquad \\
    \label{theorem-general-representer-2}
    h_{P,\lambda}(z,y)
    \!\!\!&\in&\!\!\!
    \partial L\big(z,y,(Af_{A,P,\lambda})(z)\big)
  \end{eqnarray}  
  where $\partial L(z,y,\cdot)$ denotes the subdifferential
  of the convex function $t\mapsto L(z,y,t)$.
 \item[(b)] If $A_P^\ast:\;L_2(P)\rightarrow H$ denotes the adjoint
  of the continuous linear map
  $A_P:\;H\rightarrow L_2(P)$ given by 
  $\,(A_Pf)(z,y)=(Af)(z)\;\;
   \forall\,f\in H,\;z\in\mathcal{Z},\;y\in\mathcal{Y}
  $, then
  \begin{eqnarray}\label{theorem-general-representer-3}
    f_{A,P,\lambda}
    &=&-\frac{1}{2\lambda}A_P^\ast(h_{P,\lambda})\qquad\qquad
  \end{eqnarray}
  and
  \begin{eqnarray}\label{theorem-general-representer-4}
    f_{A,P,\lambda}(x)\!\!\!
    &=&\!\!\!
       -\frac{1}{2\lambda}\int\!\!A\big(\Phi(x)\big)(z)h_{P,\lambda}(z,y)\,
                          P\big(d(z,y)\big)
        \quad\forall\,x\in\mathcal{X}.\qquad
  \end{eqnarray}
 \item[(c)] If $P_1$ is a probability measure on 
  $\mathcal{Z}\times\mathcal{Y}$ such that $\int b\,dP_1<\infty$
  and $\int b_0^{\prime\,2}\,dP_1<\infty$,
  then 
  \begin{eqnarray}\label{theorem-general-representer-5}
    \big\|f_{A,P_1,\lambda}-f_{A,P,\lambda}\big\|_H
    &\leq&\frac{1}{\lambda}\sup_{f\in\mathcal{F}}
          \left|\int h_{P,\lambda}Af\,dP_1 
                - \int h_{P,\lambda}Af\,dP
          \right|\qquad
  \end{eqnarray}
  where $\mathcal{F}=\big\{f\in H\big|\,\|f\|_H\leq 1\big\}$.
 \end{enumerate}
\end{theorem}

The following theorem yields a rate of convergence for the stochastic part,
that is, the difference between the empirical estimate
$f_{A,\mathbf{D}_n,\lambda_n}$ and its theoretical counterpart
$f_{A,P,\lambda_n}$. In case of ordinary regularized 
kernel methods (i.e. $A=\text{id}$), a corresponding result can simply be 
proven by the representer theorem and Hoeffding's inequality for Hilbert spaces.
However, in our case of inverse problems, the situation is much more complicated
because working with property (\ref{theorem-general-representer-5}) of the 
representer theorem for inverse problems is more troublesome than working
with the corresponding property in case of $A=\text{id}$. Accordingly,
we cannot apply Hoeffding's inequality offhand but use Donsker theory for
empirical processes instead.
\begin{theorem}\label{theorem-convergence-stochastic-term}
  Let Assumptions \ref{assumption-general-setting} and \ref{assumption-on-L}
  be fulfilled.
  Let $(\lambda_n)_{n\in\mathbb{N}}\subset(0,\infty)$ and 
  $(a_n)_{n\in\mathbb{N}}\subset(0,\infty)$ be sequences
  such that
  \begin{eqnarray}\label{theorem-convergence-stochastic-term-1}
    \lim_{n\rightarrow\infty}\lambda_n=0,\quad\;
    \lim_{n\rightarrow\infty}a_n=\infty,\quad\;\text{and}\quad\;
    \lim_{n\rightarrow\infty}
    \frac{a_n}{\lambda_n^{1+p/2}\sqrt{n}\,}
    \;=\;0\,.\quad
  \end{eqnarray}
  Let (\ref{assumption-operator-A-continuous-on-Cb}) be fulfilled for $A$.
  Then,
  $$a_n\big\|f_{A,\mathbf{D}_n,\lambda_n}-f_{A,P,\lambda_n}\big\|_H
    \;\;\xrightarrow[\;n\rightarrow\infty\;]{}\;\;0
    \qquad\text{in probability}.
  $$
\end{theorem}
In the following, we are concerned with the deterministic part.
Prop.\ \ref{prop-convergence-risk-deterministic-part} states that the
risk of $f_{A,P,\lambda_n}$ converges to the infimal risk; 
Theorem \ref{theorem-convergence-deterministic-part} yields that 
$f_{A,P,\lambda_n}$ even converges in $H$-norm to a minimizer
of the risk -- provided that a minimizer exists in $H$.
\begin{proposition}\label{prop-convergence-risk-deterministic-part}
  Let Assumption \ref{assumption-general-setting} and
  (\ref{theorem-unique-existence-theoretical-svm-1}) be fulfilled, and let
  $(\lambda_n)_{n\in\mathbb{N}}\subset (0,\infty)$ be a sequence such that
  $\lim_{n\rightarrow\infty}\lambda_n=0$. Then,
  $$\lim_{n\rightarrow\infty}\mathcal{R}_{A,P}\big(f_{A,P,\lambda_n}\big)
    \;=\;\inf_{f\in H}\mathcal{R}_{A,P}(f)\,.
  $$ 
\end{proposition}
\begin{theorem}\label{theorem-convergence-deterministic-part}
  Let Assumptions \ref{assumption-general-setting} and \ref{assumption-on-L}
  be fulfilled,
  and let $A$ fulfill (\ref{assumption-operator-A-continuous-on-Cb}).
  Assume that 
  \begin{eqnarray}\label{theorem-convergence-deterministic-part-1}
    \exists\,f^\ast \in H\text{ s.t. }
    \mathcal{R}_{A,P}(f^\ast)\;=\;\inf_{f\in H}\mathcal{R}_{A,P}(f)\,.
  \end{eqnarray}
  Then, there is a unique $f_{A,P}\in H$ with the following two
  properties:
  \begin{eqnarray}\label{theorem-convergence-deterministic-part-2}
    \mathcal{R}_{A,P}(f_{A,P})\;=\;\inf_{f\in H}\mathcal{R}_{A,P}(f)\,,
  \end{eqnarray}
  \begin{eqnarray}\label{theorem-convergence-deterministic-part-3}
    f^\ast\!\in\! H,\;
    \mathcal{R}_{A,P}(f^\ast)=\!\inf_{f\in H}\mathcal{R}_{A,P}(f)
    \;\;\Rightarrow\;\;
    \|f^\ast\|_H \!>\! \|f_{A,P}\|_H\text{ or }f^\ast\!=\!f_{A,P}.\;
  \end{eqnarray}
  Furthermore, for every sequence 
  $(\lambda_n)_{n\in\mathbb{N}}\subset(0,\infty)$
  such that $\lim_{n\rightarrow\infty}\lambda_n=0$, it follows that
  \begin{eqnarray}\label{theorem-convergence-deterministic-part-4}
    \lim_{n\rightarrow\infty}\big\|f_{A,P,\lambda_n}-f_{A,P}\big\|_H
    \;=\;0\;.
  \end{eqnarray}
\end{theorem}

The following Lemma \ref{lemma-bounding-the-distance-between-functions-by-risks}
provides us with a bound of the form (\ref{form-of-a-bound-functions-and-risks}).
Such a bound could also easily be adopted from
the general results in \cite[\S\,3.9]{steinwart2008}. However, 
in our special situation, it is possible to obtain a tighter bound
which enables the proof of better rates of convergence in Theorem
\ref{theorem-rate-of-convergence}.
\begin{lemma}\label{lemma-bounding-the-distance-between-functions-by-risks}
  Let $P$ be a probability measure on $\mathcal{Z}\times\mathcal{Y}$
  such that the marginal distribution $P_{\mathcal{Y}}$ has a finite first moment,
  i.e., $\int |y|\,P(d(z,y))<\infty$.
  Let $L$ be the absolute deviation loss 
  (\ref{absolute-deviation-loss}) and assume the heteroscedastic
  model given by 
  (\ref{heteroscedastic-regression-model-nochmal})--%
  (\ref{heteroscedastic-regression-model-assumption-error-density-2}).
  Define $\alpha:=a_h\underline{c}_s>0$ 
  and $t_z^\ast:=\big(Af_{A,P}\big)(z)$
  for every $z\in\mathcal{Z}$.
  Then, there is a $B\in\mathfrak{B}_{\mathcal{Z}}$ such that
  $P_{\mathcal{Z}}(B)=1$ and, for every $z\in B$ and 
  $t\in\big(t_z^\ast-\alpha , t_z^\ast+\alpha\big)$,
  $$\frac{c_h}{2\overline{c}_s}\cdot
    \big(t-t_z^\ast\big)^2\;\leq\;
    \int L(z,y,t)\,P(dy|z)\,-\,\int L(z,y,t_z^\ast)\,P(dy|z)\,.
  $$   
\end{lemma}

\subsection{Proofs}

\begin{proof}
  \item[\textbf{Proof of Theorem 
        \ref{theorem-empirical-representer-theorem}:}
       ]
  Fix any 
  $D_n=\big((z_1,y_1),\dots,(z_n,y_n)\big)
    \in(\mathcal{Z}\times\mathcal{Y})^n
  $ 
  and $\lambda\in(0,\infty)$.
  Existence and uniqueness of $f_{A,D_n,\lambda}$
  defined by (\ref{def-empirical-svm}) follow from
  Theorem \ref{theorem-unique-existence-theoretical-svm}
  by choosing the empirical measure for $P$.
  Define
  $$\tilde{A}\,:\;\;H\;\rightarrow\;\mathbb{R}^n,\qquad
    f\;\mapsto\;
    \Big((Af)(z_1)\,,\, \hdots \,,\, (Af)(z_n)\Big)^{\mathsf{T}}\;.
  $$
  The assumptions on $A$ imply that $\tilde{A}$ is again linear and
  continuous. Let $\tilde{A}^\ast$ denote the adjoint operator of
  $\tilde{A}$.
  Then, for every $x\in\mathcal{X}$,
  \begin{eqnarray}\label{theorem-empirical-representer-theorem-p1}
    \big(\tilde{A}^\ast(e_i)\big)(x)
    =\big\langle \tilde{A}^\ast(e_i),\Phi(x)\big\rangle_H
       =\big\langle e_i,\tilde{A}\big(\Phi(x)\big)
            \big\rangle_{\mathbb{R}^n}
    =\Big(A\big(\Phi(x)\big)\Big)(z_i)\;\;
  \end{eqnarray}
  and, therefore,
  \begin{eqnarray}\label{theorem-empirical-representer-theorem-p2}
    M_{i,j}\!\!\!
    &=&\!\!\!
       \bigg(\!A\Big(\big(A\Phi(\cdot)\big)(z_i)\Big)\!\bigg)(z_j)
       \,\stackrel{(\ref{theorem-empirical-representer-theorem-p1})}{=}\,
          \Big(\!A\big(\tilde{A}^\ast(e_i)\big)\!\Big)(z_j)
       \,=\,\Big\langle e_j,\tilde{A}\big(\tilde{A}^\ast(e_i)\big)
            \Big\rangle_{\mathbb{R}^n}\;\; \nonumber \\
    &=&\!\!\!
       \big\langle \tilde{A}^\ast(e_j),\tilde{A}^\ast(e_i)
       \big\rangle_{H}
  \end{eqnarray}
  This implies that the matrix $M$ is symmetric and, in addition, that
  it is positive semi-definite because, for every
  $\alpha=(\alpha_1,\dots,\alpha_n)^{\mathsf{T}}\in\mathbb{R}^n$,
  \begin{eqnarray}\label{theorem-empirical-representer-theorem-p3}
    \alpha^{\mathsf{T}}M\alpha
    \;\stackrel{(\ref{theorem-empirical-representer-theorem-p2})}{=}\;
       \sum_{i,j}\alpha_i\alpha_j
       \big\langle \tilde{A}^\ast(e_i),\tilde{A}^\ast(e_j)
       \big\rangle_{H}
    \;=\;\bigg\|\sum_i \alpha_i\tilde{A}^\ast(e_i)\bigg\|_{H}^2
    \;\geq\;0\,.
  \end{eqnarray}
  Fix any $\alpha=(\alpha_1,\dots,\alpha_n)^{\mathsf{T}}\in\mathbb{R}^n$
  and define
  \begin{eqnarray}\label{theorem-empirical-representer-theorem-p6}
    f_0(x)\;=\;\sum_{i=1}^n\alpha_i\!\cdot\!\big(A\Phi(x)\big)(z_i)
    \qquad\forall\,x\in\mathcal{X}\,.
  \end{eqnarray}
  Note that (\ref{theorem-empirical-representer-theorem-p1}) implies
  that $f_0\in H$ and
  \begin{eqnarray}\label{theorem-empirical-representer-theorem-p4}
    \|f_0\|_H^2
    \;\stackrel{(\ref{theorem-empirical-representer-theorem-p1})}{=}\;
    \bigg\|\sum_i \alpha_i\tilde{A}^\ast(e_i)\bigg\|_{H}^2
    \;\stackrel{(\ref{theorem-empirical-representer-theorem-p3})}{=}\;
    \alpha^{\mathsf{T}}M\alpha
  \end{eqnarray}
  Furthermore,
  \begin{eqnarray}\label{theorem-empirical-representer-theorem-p5}
    \big(Af_0\big)(z_j)
    &=&
       \big\langle e_j,\tilde{A}(f_0)
       \big\rangle_{\mathbb{R}^n}
    \;\stackrel{(\ref{theorem-empirical-representer-theorem-p6},
                 \ref{theorem-empirical-representer-theorem-p1}
                )
               }{=}\;
        \sum_{i=1}^n\alpha_i
           \Big\langle e_j, \tilde{A}\big(\tilde{A}^\ast(e_i)\big)
           \Big\rangle_{\mathbb{R}^n}\;= \nonumber \\
    &\stackrel{(\ref{theorem-empirical-representer-theorem-p2})}{=}&
        \sum_{i=1}^n\alpha_i M_{i,j}
        \;=\;\alpha^{\mathsf{T}}Me_j\;.
  \end{eqnarray}
  Hence, (\ref{theorem-empirical-representer-theorem-1}) follows from 
  the definition of the regularized risk $\mathcal{R}_{A,D,\lambda}$,
  (\ref{theorem-empirical-representer-theorem-p4}), and
  (\ref{theorem-empirical-representer-theorem-p5}). \\
  It only remains to prove (\ref{theorem-empirical-representer-theorem-2}),
  which can be done
  similarly to the proof of 
  \cite[Theorem 3.2]{krebs:louis:2009} and \cite[Lemma 3.1]{krebs2011}.
  The main idea of the proof is to show that  $f_{A,D_n,\lambda}$ is an 
  element of the image $\text{im}\big(\tilde{A}^\ast\big)$. 
  First, note that the image 
  $\text{im}\big(\tilde{A}^\ast\big)$ is a 
  finite-dimensional linear subspace of
  $H$, hence, it is closed; see, e.g.\ 
  \cite[Cor.\ 3.2.17]{denkowski2003}.
  Then, for every $f_0\in H$, there is an
  $f_1\in\text{im}\big(\tilde{A}^\ast\big)$ and an
  $f_2\in\big(\text{im}\big(\tilde{A}^\ast\big)\big)^{\bot}$ such that
  $f_0=f_1+f_2$; see, e.g., 
  \cite[Cor.\ 3.7.16]{denkowski2003}. Then,
  \begin{eqnarray*}
    \big(Af_0)(z_i)
    &=&\big(Af_1)(z_i)+\big(Af_2)(z_i)
       \;=\;\big(Af_1)(z_i)+\big\langle e_i,\tilde{A}(f_2)\big\rangle_H\;=\\
    &=&\big(Af_1)(z_i)+\big\langle \tilde{A}^\ast(e_i),f_2\big\rangle_H
       \;=\;\big(Af_1)(z_i)
  \end{eqnarray*}
  and
  $$\|f_0\|_H^2\;=\;\|f_1\|_H^2+\|f_2\|_H^2\;.
  $$
  This shows that, if $f_0$ is not in the image of $\tilde{A}^\ast$,
  then there is another $f_1\in H$ such that
  $\mathcal{R}_{A,D,\lambda}(f_1)<\mathcal{R}_{A,D,\lambda}(f_0)$.
  Hence, the minimizer $f_{A,D_n,\lambda}$ is
  in the image of $\tilde{A}^\ast$, that is, there are 
  $\alpha_1,\dots,\alpha_n\in\mathbb{R}$ such that
  $$f_{A,D_n,\lambda}\;=\;
    \tilde{A}^\ast\bigg(\sum_{i=1}^n\alpha_i e_i\bigg)\;=\;
    \sum_{i=1}^n\alpha_i\tilde{A}^\ast(e_i)
    \;\stackrel{(\ref{theorem-empirical-representer-theorem-p1})}{=}\;
    \sum_{i=1}^n\alpha_i\cdot\big(A\Phi(\cdot)\big)(z_i)\;.
  $$
\end{proof}

\begin{proof}
  \item[\textbf{Proof of Theorem 
        \ref{theorem-unique-existence-theoretical-svm}:}
       ]
    Consider $\mathcal{R}_{A,P,\lambda}$ as a map from $H$ to
    $\mathbb{R}\cup\{\pm\infty\}$. the map $\mathcal{R}_{A,P,\lambda}$
    is convex and fulfills 
    $\mathcal{R}_{A,P,\lambda}(f)\geq 0>-\infty$ for every $f\in H$ and
    $\lim_{\|f\|_H\rightarrow\infty}\mathcal{R}_{A,P,\lambda}(f)=\infty$
    because
    $$\liminf_{\|f\|_H\rightarrow\infty}\mathcal{R}_{A,P,\lambda}(f)
      \;\geq\;\lambda\liminf_{\|f\|_H\rightarrow\infty}\|f\|_H^2
      \;=\;\infty\;.
    $$  
    According to Assumption (\ref{assumption-operator-A-continuous-on-H}),
    $$f_n\;\xrightarrow[\;n\rightarrow\infty\;]{}\;f \quad\text{in }H
      \qquad\Rightarrow\qquad
      (Af_n)(z)\;\xrightarrow[\;n\rightarrow\infty\;]{}\;(Af)(z)
      \quad\forall\,z\in\mathcal{Z}\;.
    $$
    so that  
    Fatou's Lemma -- e.g.\ 
    \cite[Theorem 2.2.17]{denkowski2003} -- implies that
    $\mathcal{R}_{A,P,\lambda}$ is lower semicontinuous.
    Then, it follows from 
    \cite[Prop.\ 5.2.12]{denkowski2003} that there is an
    $f_{A,P,\lambda}\in H$ which minimizes 
    $\mathcal{R}_{A,P,\lambda}$ in $H$. Assumption 
    (\ref{theorem-unique-existence-theoretical-svm-1}) 
    implies $\mathcal{R}_{A,P}(f_{A,P,\lambda})<\infty$ so that
    uniqueness of $f_{A,P,\lambda}\in H$
    follows from strict convexity of the squared norm and
    convexity of $\mathcal{R}_{A,P}$.
\end{proof}

\begin{proof}
  \item[\textbf{Proof of Prop.\ 
        \ref{prop-compactness-A}:}
       ]
  See, e.g., \cite[Prop.\ 3.7.47]{denkowski2003} for the fact that
  compactness of $A:H\rightarrow\mathcal{C}_b(\mathcal{Z})$ is equivalent to
  (\ref{prop-compactness-A-1}). In order to show (\ref{prop-compactness-A-1}),
  fix any sequence $(f_n)_{n\in \mathbb{N}}\subset H$ which converges weakly
  in $H$ to some $f_0\in H$ for $n\rightarrow\infty$.
  As a weakly convergent sequence in a Hilbert space is bounded
  \cite[see, e.g.,][Cor.\ 3.4.10]{denkowski2003}, 
  there is a 
  $c\in(0,\infty)$ such that, for every $n\in\mathbb{N}$,
  we have $\|f_n\|_H\leq c$.
  Hence, for every sequence $x_\ell\rightarrow x_0$ in $\mathcal{X}$,
  \begin{eqnarray*}
    \lefteqn{
    \lim_{\ell\rightarrow\infty}\sup_{n\in\mathbb{N}}\big|f_n(x_\ell)-f_n(x_0)\big|
    \;\stackrel{(\ref{reproducing-property})}{=}\;
    \lim_{\ell\rightarrow\infty}\sup_{n\in\mathbb{N}}
        \big|\big\langle f_n,\Phi(x_\ell)-\Phi(x_0)\big\rangle_H\big|\;\leq
    }\\
    &\leq&\!\!
          \lim_{\ell\rightarrow\infty}\sup_{n\in\mathbb{N}} 
               \big\|f_n\big\|_H\!\cdot\!\big\|\Phi(x_\ell)-\Phi(x_0)\big\|_H
          \,\leq\,c\!\cdot\!\lim_{\ell\rightarrow\infty}
                           \big\|\Phi(x_\ell)-\Phi(x_0)\big\|_H 
          =\,0
  \end{eqnarray*}
  since continuity of $k$ implies continuity of $\Phi$; see, e.g.,
  \cite[Lemma 4.29]{steinwart2008}. That is, we have shown that
  the sequence $(f_n)_{n\in \mathbb{N}}$
  is equicontinuous. In addition, it follows from 
  weak convergence in $H$ and  
  the reproducing property (\ref{reproducing-property}) 
  that $f_n$ converges to $f_0$ pointwise.
  Since $\mathcal{X}$ is compact, pointwise
  convergence together with equicontinuity implies uniform convergence
  of $f_n$ to $f_0$; see, e.g., 
  \cite[Prop.\ 1.6.14 and Theorem 1.6.12]{denkowski2003}.
  Hence, the statement follows from assumption 
  (\ref{assumption-operator-A-continuous-on-Cb}).
\end{proof}

\begin{proof}
  \item[\textbf{Proof of Theorem 
        \ref{theorem-general-representer}:}
       ]
  We start with the proof of (a) and (b).     
  Define 
  $$\mathcal{Q}_P\;:\;\;L_2(P)\;\rightarrow\;\mathbb{R}\,,\qquad
    g\;\mapsto\;\int L\big(z,y,g(z,y)\big)\,P\big(d(z,y)\big)\;.
  $$ 
  That is, the risk $\mathcal{R}_{A,P}:\;H\rightarrow\mathbb{R}$
  is given by $\mathcal{R}_{A,P}(f)=\big(\mathcal{Q}_P\circ A_P\big)(f)$,
  $f\in H$.
  Note that Assumption \ref{assumption-on-L} implies that
  $\mathcal{Q}_P$ is defined well.
  Then, it follows from 
  \cite[Prop.\ 2C and Cor.\ 3E]{Rockafellar:1976} that
  the subdifferential of the convex map $\mathcal{Q}_P$ is given by
  $$\partial\mathcal{Q}_P(g)\,=\,
    \Big\{h\in L_2(P)\;
    \Big|\;\;h(z,y)\in\partial L\big(z,y,g(z,y)\big)
             \quad P\big(d(z,y)\big)\,-\,\text{a.s.}
    \Big\};
  $$
  see, e.g., \cite[Prop.\ A.6.13]{steinwart2008}. 
  It is easy to see from Assumption \ref{assumption-on-L} that
  $\mathcal{Q}_P$ is continuous on $L_2(P)$ and, therefore,  
  $$\partial\mathcal{R}_{A,P}(f)\;=\;
    \partial\big(\mathcal{Q}_P\circ A_P\big)(f)\;=\;
    A_P^\ast\big(\partial\mathcal{Q}_P(A_Pf)\big)\;;
  $$
  see, e.g., \cite[Theorem 5.3.33]{denkowski2003}.
  That is,
  $$\partial\mathcal{R}_{A,P}(f)=
    \bigg\{A_P^\ast(h)\,
    \bigg|
           \begin{array}{l}
             \qquad h\in L_2(P),\\
             h(z,y)\in\partial L\big(z,y,\big(A_Pf\big)(z,y)\big)
             \;\; P\big(d(z,y)\big)\text{\,-\,a.s.}
           \end{array}\!\!\!
    \bigg\}.
  $$
  The convex map $H\rightarrow\mathbb{R}$, $f\mapsto\lambda\|f\|_H^2$
  is Fr\'{e}chet differentiable with derivative $2\lambda f$
  -- see, e.g., \cite[Example 5.1.6(c)]{denkowski2003}
  -- and, therefore, its subdifferential at $f\in H$ is given by
  $\{2\lambda f\}$ -- see, e.g.,
  \cite[Prop.\ 5.3.30]{denkowski2003}. Since
  $\mathcal{R}_{A,P,\lambda}(f)=\mathcal{R}_{A,P}(f)+\lambda\|f\|_H^2$
  for every $f\in H$, it follows that
  $$\partial\mathcal{R}_{A,P,\lambda}(f)\;=\;
    2\lambda f + \partial\mathcal{R}_{A,P}(f)\,,
    \qquad f\in H\,; 
  $$
  see, e.g., \cite[Theorem 5.3.32]{denkowski2003}.
  Since $\mathcal{R}_{A,P,\lambda}$ attains its minimum in $H$ at
  $f_{A,P,\lambda}$, it follows from the definition of the subdifferential
  that $0\in\partial\mathcal{R}_{A,P,\lambda}(f_{A,P,\lambda})$.
  That is, there is an $h\in\mathcal{L}_2(P)$ such that
  $h(z,y)\in\partial L\big(z,y,\big(A_Pf\big)(z,y)\big)$
  for $P$\,-\,a.e. $(z,y)\in\mathcal{Z}\times\mathcal{Y}$ and
  $f_{A,P,\lambda}=-\frac{1}{2\lambda}A_P^\ast(h)$.
  In the following, it is shown that we can even choose 
  $h_{P,\lambda}\in\mathcal{L}_2(P)$ such that
  $h_{P,\lambda}(z,y)\in\partial L\big(z,y,\big(A_Pf\big)(z,y)\big)$
  for \textit{every} $(z,y)\in\mathcal{Z}\times\mathcal{Y}$. For every
  $(z,y)\in\mathcal{Z}\times\mathcal{Y}$, let $L_+^\prime(z,y,\cdot)$
  denote the right derivative function of $L(z,y,\cdot)$.
  Recall from \cite[Theorem 24.1 and p.\ 229]{Rockafellar:1970}, that 
  this is a function 
  $L_+^\prime(z,y,\cdot):\mathbb{R}\rightarrow\mathbb{R}$
  and $L_+^\prime(z,y,t)\in\partial L(z,y,t)$ for every
  $(z,y)\in\mathcal{Z}\times\mathcal{Y}$ and $t\in\mathbb{R}$.
  Since the function $L_+^\prime:\;(z,y,t)\mapsto L_+^\prime(z,y,t)$
  is the pointwise limit of a sequence of measurable functions,
  $L_+^\prime$ is measurable. Hence, there is a $P$-null-set 
  $N\in\mathfrak{B}_{\mathcal{Z}\times\mathcal{Y}}$ such that
  $h_{P,\lambda}\in\mathcal{L}_2(P)$ defined by
  $$h_{P,\lambda}(z,y)\;=\;
    h(z,y)I_{N^\complement}(z,y)
    +L_+^\prime\big(z,y,\big(A_Pf\big)(z,y)\big)I_{N}(z,y)
  $$
  fulfills (\ref{theorem-general-representer-2}) and
  (\ref{theorem-general-representer-3}). Next, 
  (\ref{theorem-general-representer-4}) follows from
  \begin{eqnarray*}
    f_{A,P,\lambda}(x)
    &\stackrel{(\ref{reproducing-property})}{=}&
       \big\langle f_{A,P,\lambda},\Phi(x) \big\rangle_H
       \;\stackrel{(\ref{theorem-general-representer-3})}{=}\;
       -\frac{1}{2\lambda}
         \big\langle A_P^\ast h_{P,\lambda},\Phi(x) \big\rangle_H \;=\\
    &=&-\frac{1}{2\lambda}
         \big\langle h_{P,\lambda}, A_P\big(\Phi(x)\big) 
         \big\rangle_{L_2(P)}
       \;=\;-\frac{1}{2\lambda}\int\!\! A\big(\Phi(x)\big)h_{P,\lambda}\,
                          dP \quad
  \end{eqnarray*}
  for every $x\in\mathcal{X}$. Finally, in order to prove
  (\ref{theorem-general-representer-1}), note that
  $$\lambda\big\|f_{A,P,\lambda}\big\|_H^2\;\leq\;
    \mathcal{R}_{A,P,\lambda}\big(f_{A,P,\lambda}\big)\;\leq\;
    \mathcal{R}_{A,P,\lambda}\big(0\big)
    \;\stackrel{(\ref{assumption-on-L-1})}{\leq}\;\int b\,dP
  $$
  and, therefore, 
  (\ref{assumption-operator-A-continuous-on-H}) implies
  \begin{eqnarray}\label{theorem-general-representer-p1}
    \big\|Af_{A,P,\lambda}\big\|_\infty\,\leq\,\|A\|\!\cdot\!
           \big\|f_{A,P,\lambda}\big\|_H
     \,<\,\|A\|\sqrt{\frac{1}{\lambda}\int b\,dP\,}+1
     \;=:\;a\;<\infty.\quad
  \end{eqnarray}
  Now, fix any $(z,y)\in\mathcal{Z}\times\mathcal{Y}$,
  $t\in(-a,a)$, and $\gamma_t\in\partial L(z,y,t)$. 
  Then, the definition of the subdifferential implies
  $$\gamma_t\cdot(a-t)
    \;\leq\;L(z,y,a)-L(z,y,t)
    \;\stackrel{(\ref{assumption-on-L-2})}{\leq}\;
    \big(b_0^\prime(y)+b_1^\prime a^p\big)\cdot|a-t|
  $$
  and
  $$\gamma_t\cdot(-a-t)
    \;\leq\;L(z,y,-a)-L(z,y,t)
    \;\stackrel{(\ref{assumption-on-L-2})}{\leq}\;
    \big(b_0^\prime(y)+b_1^\prime a^p\big)\cdot|-a-t|\;.
  $$
  Dividing these inequalities by $|a-t|=a-t$ and $-|\!-a-t|=-a-t$, 
  respectively, leads to
  \begin{eqnarray}\label{theorem-general-representer-p2}
    \big|\gamma_t\big|\;\leq\;
    b_0^\prime(y)+b_1^\prime a^p\;.
  \end{eqnarray}
  Due to (\ref{theorem-general-representer-p1}), we may choose
  $t=(Af_{A,P,\lambda})(z)\in(-a,a)$ so that
  (\ref{theorem-general-representer-2}) and
  (\ref{theorem-general-representer-p2}) yield
  $$\big|h_{P,\lambda}(z,y)\big|\;\leq\;
    b_0^\prime(y)+b_1^\prime a^p\;.
  $$  
  Then, (\ref{theorem-general-representer-1}) follows from
  the definition of $a$ in 
  (\ref{theorem-general-representer-p1}); that is, we have proven 
  parts (a) and (b) of the theorem. \\
  For the proof of part (c), let
  $P_1$ be any probability measure on 
  $\mathcal{Z}\times\mathcal{Y}$ such that $\int b\,dP_1<\infty$
  and $\int b_0^{\prime\,2}\,dP_1<\infty$.
  In order to shorten the notation, define
  $h_0:=h_{P,\lambda}$, $f_0:=f_{A,P,\lambda}$, and
  $f_1:=f_{A,P_1,\lambda}$. Then, (\ref{theorem-general-representer-2})
  implies
  \begin{eqnarray}\label{theorem-general-representer-p3}
    h_0(z,y)\big(Af_1(z)-Af_0(z)\big)\;\leq\;
    L\big(z,y,Af_1(z)\big)-L\big(z,y,Af_0(z)\big)
  \end{eqnarray}
  for every $(z,y)\in\mathcal{Z}\times\mathcal{Y}$.
  The map $A_{P_1}:H\rightarrow L_2(P_1)$ defined by
  $\big(A_{P_1}f\big)(z,y)=\big(Af\big)(z)$ is a continuous linear
  operator; let $A_{P_1}^\ast$ denote its adjoint operator. 
  Since $h_0\in L_2(P_1)$ according to 
  (\ref{theorem-general-representer-1}) and the assumptions on
  $P_1$, it follows that
  \begin{eqnarray*}
    \int h_0(z,y)\big(Af_1(z)-Af_0(z)\big)\,P_1\big(d(z,y)\big)
    &=&\big\langle h_0, A_{P_1}(f_1-f_0)\big\rangle_{L_2(P_1)}
       \;=\\
    &=&\big\langle f_1-f_0,A_{P_1}^\ast h_0\big\rangle_H\;.
  \end{eqnarray*}
  Hence, integrating both
  sides of (\ref{theorem-general-representer-p3}) 
  with respect to $P_1$ implies
  \begin{eqnarray}\label{theorem-general-representer-p4}
    \big\langle f_1-f_0,A_{P_1}^\ast h_0\big\rangle_H
    &\leq&\!\!\mathcal{R}_{A,P_1}(f_1)-\mathcal{R}_{A,P_1}(f_0)\,.
    \qquad\qquad
  \end{eqnarray}
  An elementary calculation shows
  \begin{eqnarray}\label{theorem-general-representer-p5}
    2\lambda\big\langle f_1-f_0,f_0\big\rangle_H
    +\lambda\big\|f_0-f_1\big\|_H^2
    &=&\lambda\big\|f_1\big\|_H^2-\lambda\big\|f_0\big\|_H^2\,.
    \qquad\qquad
  \end{eqnarray} 
  Then, calculating 
  $(\ref{theorem-general-representer-p4})
   +(\ref{theorem-general-representer-p5})
  $,
  the definition of the regularized risk 
  $\mathcal{R}_{A,P_1,\lambda}$, and the definition of 
  $f_1=f_{A,P_1,\lambda}$ imply
  $$\big\langle f_1-f_0,A_{P_1}^\ast h_0+2\lambda f_0\big\rangle_{\!H}
    +\lambda\big\|f_0-f_1\big\|_H^2
    \,\leq\,\mathcal{R}_{A,P_1,\lambda}(f_1)
             -\mathcal{R}_{A,P_1,\lambda}(f_0)
    \,\leq\, 0.
  $$
  Hence, it follows from (\ref{theorem-general-representer-3}) that
  $$\lambda\big\|f_0-f_1\big\|_H^2
    \,\leq\,
    \big\langle f_0-f_1,A_{P_1}^\ast h_0-A_{P}^\ast h_0\big\rangle_{\!H}
    \,\leq\,
    \big\|f_0-f_1\big\|_H\big\|A_{P_1}^\ast h_0-A_{P}^\ast h_0\big\|_H
  $$
  and, therefore,
  \begin{eqnarray*}
    \lefteqn{
    \lambda\big\|f_1-f_0\big\|_H
    \,\leq\,\big\|A_{P_1}^\ast h_0-A_{P}^\ast h_0\big\|_H\,=
          \sup_{f\in \mathcal{F}}\!
          \big|\big\langle f,A_{P_1}^\ast h_0 \big\rangle_H
               -\big\langle f,A_{P}^\ast h_0 \big\rangle_H
          \big|
    }\\
    &=&\!\!\!
          \sup_{f\in \mathcal{F}}
          \Big|\big\langle A_{P_1}f, h_0 \big\rangle_{L_2(P_1)}
               -\big\langle A_{P}f, h_0 \big\rangle_{L_2(P)}
          \Big|\;=\\
    &=&\!\!\!
          \sup_{f\in \mathcal{F}}
          \left|
            \int\! h_0(z,y)\big(Af\big)(z)\,P_1\big(d(z,y)\big)
            -\!\!\int\! h_0(z,y)\big(Af\big)(z)\,P\big(d(z,y)\big)
          \right| .
  \end{eqnarray*}
\end{proof}

\begin{proof}
  \item[\textbf{Proof of Theorem 
        \ref{theorem-convergence-stochastic-term}:}
       ]  
  For every $n\in\mathbb{N}$, take $h_n:=h_{P,\lambda_n}$
  from Theorem \ref{theorem-general-representer} and define
  $$g_{n,f}\;:=\;\frac{a_n}{\lambda_n\sqrt{n}\,}h_n Af
    \qquad\forall\,f\in \mathcal{F}
  $$
  where $\mathcal{F}=\big\{f\in H\big|\,\|f\|_H\leq 1\big\}$.
  Then,  
  $$\mathcal{G}_n\;:=\;
    \big\{g_{n,f}\;\big|\;f\in\mathcal{F}\big\}
  $$
  is a changing class in the sense of
  \cite[\S\,19.5]{vandervaart1998} and we can use a Donsker theorem
  for such classes in order to prove
  \begin{eqnarray}\label{theorem-convergence-stochastic-term-p1}
    \sqrt{n}\left(\frac{1}{n}\sum_{i=1}^n g_{n,f}(Z_i,Y_i)
                  -\mathbb{E}_P g_{n,f}
            \right)_{f\in\mathcal{F}}
    \;\;\leadsto\;\;0
    \qquad\text{in}\;\;\;\ell_{\infty}(\mathcal{F})\quad
  \end{eqnarray}   
  below. Then, it follows from 
  (\ref{theorem-convergence-stochastic-term-p1}), the definition of 
  $g_{n,f}$, and the continuous mapping theorem 
  \cite[e.g.,][Theorem 1.3.6]{vandervaartwellner1996} that
  \begin{eqnarray}\label{theorem-convergence-stochastic-term-p2}
    \frac{a_n}{\lambda_n}\sup_{f\in\mathcal{F}}
          \left|\frac{1}{n}\sum_{i=1}^n 
                             h_{P,\lambda_n}(Z_i,Y_i)Af(Z_i) 
                - \int h_{P,\lambda_n}Af\,dP
          \right|
    \;\;\leadsto\;\;0\,.
    \qquad
  \end{eqnarray}
  Since weak convergence to a constant implies convergence in
  (outer) probability
  \cite[see, e.g.,][Lemma 1.10.2]{vandervaartwellner1996}, the statement of
  Theorem \ref{theorem-convergence-stochastic-term} follows from
  (\ref{theorem-convergence-stochastic-term-p2}) and  
  (\ref{theorem-general-representer-5}). 
  Measurability of the random variable 
  $\|f_{A,\mathbf{D}_n,\lambda_n}-f_{A,P,\lambda_n}\|_H$
  can be proven by simply following the lines of the proof of
  \cite[Lemma 9a]{hable2013}. (The only noteworthy difference is that
  proving the analogue of \cite[(54)]{hable2013} involves an application
  of Prop.\ \ref{prop-compactness-A}.) \\
  Hence, it only remains to
  prove (\ref{theorem-convergence-stochastic-term-p1}) 
  and this is done by use of \cite[\S\,19.5]{vandervaart1998}
  in the following.
  To this end, note that 
  $\mathcal{G}_n=\big\{g_{n,f}\;\big|\;f\in\mathcal{F}\big\}$
  is a class of measurable functions indexed by $\mathcal{F}$.
  For the metric $\rho$ defined by $\rho(f_1,f_2)=\|f_1-f_2\|_\infty$,
  $f_1,f_2\in \mathcal{F}$, the index set $\mathcal{F}$ is
  totally bounded because $\mathcal{F}$ is relatively compact
  in $\mathcal{C}_b(\mathcal{X})$
  according to \cite[Cor.\ 4.31]{steinwart2008}.
  It follows 
  from $\lim_{n\rightarrow\infty}\lambda_n=0$ and 
  (\ref{theorem-general-representer-1})  that
  there is a $c\in(0,\infty)$ such that, for every $n\in\mathbb{N}$,
  \begin{eqnarray}\label{theorem-convergence-stochastic-term-p3}
    \big|h_n(z,y)\big|\;\leq\;b_0^\prime(y)+c\lambda_n^{-p/2}
    \qquad\forall\,(z,y)\in\mathcal{Z}\times\mathcal{Y}\;.
  \end{eqnarray}
  Since 
  $\big\|Af_1-Af_2\big\|_\infty\leq\,\|A\|\!\cdot\!\|f_1-f_2\|_H
   \leq\,2\|A\|
  $
  for every $f_1,f_2\in\mathcal{F}$, it follows from the definitions
  that
  \begin{eqnarray}\label{theorem-convergence-stochastic-term-p4}
    \int \big(g_{n,f_1}-g_{n,f_2}\big)^2 dP
    \;\leq\;4\|A\|^2\!\cdot\!
            \left(\frac{a_n}{\lambda_n\sqrt{n}}\right)^2
            \!\int \!h_n^2\,dP
    \qquad\forall\,f_1,f_2\in\mathcal{F}.\quad
  \end{eqnarray}
  Then, an easy calculation using 
  (\ref{theorem-convergence-stochastic-term-1}),
  (\ref{theorem-convergence-stochastic-term-p3}), and
  (\ref{theorem-convergence-stochastic-term-p4}) shows
  \begin{eqnarray}\label{theorem-convergence-stochastic-term-p5}
    \lim_{n\rightarrow\infty}
    \sup_{f_1,f_2\in\mathcal{F}}
    \int \big(g_{n,f_1}-g_{n,f_2}\big)^2 dP
    \;=\;0\,.
  \end{eqnarray}
  Due to (\ref{theorem-convergence-stochastic-term-1}), there is 
  a $G\in\mathcal{L}_2(P_{\mathcal{Y}})$ such that, for every 
  $n\in\mathbb{N}$ and $(z,y)\in\mathcal{Z}\times\mathcal{Y}$,
  \begin{eqnarray}\label{theorem-convergence-stochastic-term-p6}
    \frac{a_n\|A\|}{\lambda_n\sqrt{n}}\big|h_{n}(z,y)\big|
    \;\stackrel{(\ref{theorem-convergence-stochastic-term-p3})}{\leq}\;
    \frac{a_n\|A\|}{\lambda_n\sqrt{n}}b_0^\prime(y)+
    \frac{a_n\|A\|}{\lambda_n^{1+p/2}\sqrt{n}}c
    \;\,\leq\,\;G(y)\,.
  \end{eqnarray}
  Since $\|Af\|_\infty\leq\,\|A\|\!\cdot\!\|f\|_H\leq\,\|A\|$
  for every $f\in\mathcal{F}$, it follows from
  (\ref{theorem-convergence-stochastic-term-p6}) that
  \begin{eqnarray}\label{theorem-convergence-stochastic-term-p7}
    \big|g_{n,f}(z,y)\big|\;\leq\;G(y)
    \qquad\forall\,(z,y)\in\mathcal{Z}\times\mathcal{Y},\;\;
    n\in\mathbb{N},\;\; f\in\mathcal{F}.
  \end{eqnarray}
  Hence, $G$ is an envelope function of $\mathcal{G}_n$ for every
  $n\in\mathbb{N}$ which fulfills the Lindeberg conditions
  $\int G^2\,dP<\infty$ and
  $\lim_{n\rightarrow\infty}\int G^2 I_{G>\varepsilon\sqrt{n}}\,dP=0$
  for every $\varepsilon>0$.\\
  Let $\|A\|_\infty$ denote the operator norm of $A$ as an operator
  from $\mathcal{C}_b(\mathcal{X})$ to
  $\mathcal{C}_b(\mathcal{Z})$.
  Then, it follows from 
  (\ref{theorem-convergence-stochastic-term-p6}) that
  $$\big|g_{n,f_1}(z,y)-g_{n,f_2}(z,y)\big|\;\leq\;
    \|A\|_\infty G(y)\!\cdot\!\|f_1-f_2\|_\infty\;.
  $$
  According to \cite[\S\,2.7.4]{vandervaartwellner1996},
  this implies
  \begin{eqnarray}\label{theorem-convergence-stochastic-term-p8}
    N_{[\,]}\big(2\varepsilon\|A\|_\infty\|G\|_{L_2(P)},
                 \mathcal{G}_n,\|\cdot\|_{L_2(P)}
            \big)
    \;\leq\;N(\varepsilon,\mathcal{F},\|\cdot\|_\infty)
  \end{eqnarray}
  where $N_{[\,]}(\cdot)$ denotes the bracketing number
  and $N(\cdot)$ denotes the covering number; see, e.g.,
  \cite[\S\,2.1.1]{vandervaartwellner1996}. 
  According to the assumptions on $\mathcal{X}$ and
  $k$, it follows from \cite[Theorem 2.7.1]{vandervaartwellner1996}
  that there is some $C_0\in(0,\infty)$ such that, for every
  $\varepsilon>0$, we have
  $\log N(\varepsilon,\mathcal{F},\|\cdot\|_\infty)\leq
   C_0\varepsilon^{-d/m}
  $; see, e.g.,
  \cite[Eqn. (61)]{hable2012a}. Hence, it follows from
  (\ref{theorem-convergence-stochastic-term-p8}) that
  $$\log N_{[\,]}\big(\varepsilon,
                 \mathcal{G}_n,\|\cdot\|_{L_2(P)}
            \big)
    \;\leq\;C_0\!\cdot\!\big(2\|A\|_\infty\|G\|_{L_2(P)}\big)^{\frac{d}{m}}
            \cdot\varepsilon^{-\frac{d}{m}}
    \qquad\forall\,\varepsilon>0.\;
  $$
  Recall that $m>d/2$.
  Hence, the bracketing integral fulfills
  \begin{eqnarray}\label{theorem-convergence-stochastic-term-p9}
    J_{[\,]}\big(\delta_n,
                 \mathcal{G}_n,\|\cdot\|_{L_2(P)}
            \big)
    \;=\int_0^{\delta_n}\!\!\!
           \sqrt{\log N_{[\,]}\big(\varepsilon,\mathcal{G}_n,
                                   \|\cdot\|_{L_2(P)}
                              \big)
                }
         \,\,d\varepsilon
    \;\xrightarrow[\;n\rightarrow\infty\;]{}\;0\;\;
  \end{eqnarray}
  for every sequence of real numbers $\delta_n\searrow 0$.\\
  Finally, the definition of $g_{n,f}$, 
  (\ref{theorem-convergence-stochastic-term-p3}), and
  (\ref{theorem-convergence-stochastic-term-1}) imply
  $\lim_{n\rightarrow\infty}g_{n,f}(z,y)=0$
  for every $f\in\mathcal{F}$ and
  $(z,y)\in\mathcal{Z}\times\mathcal{Y}$ so that
  the dominated convergence theorem and
  (\ref{theorem-convergence-stochastic-term-p7}) yield
  \begin{eqnarray}\label{theorem-convergence-stochastic-term-p10}
    \lim_{n\rightarrow\infty}
    \mathbb{E}\big[g_{n,f_1}(Z_i,Y_i)g_{n,f_2}(Z_i,Y_i)\big]
    -\mathbb{E}g_{n,f_1}(Z_i,Y_i)
     \mathbb{E}g_{n,f_2}(Z_i,Y_i)
    \;=\;0\quad
  \end{eqnarray}
  for all $f_1,f_2\in\mathcal{F}$.
  Summing up, because of (\ref{theorem-convergence-stochastic-term-p5}),
  (\ref{theorem-convergence-stochastic-term-p7}), 
  (\ref{theorem-convergence-stochastic-term-p9}), and
  (\ref{theorem-convergence-stochastic-term-p10}), the assumptions
  in \cite[\S\,19.5]{vandervaart1998} are fulfilled and it follows
  from \cite[Theorem 19.28]{vandervaart1998} that
  $$\sqrt{n}\left(\frac{1}{n}\sum_{i=1}^n g_{n,f}(Z_i,Y_i)
                  -\mathbb{E}_P g_{n,f}
            \right)_{f\in\mathcal{F}}
    \;\;\leadsto\;\;\mathbb{G}_P
    \qquad\text{in}\;\;\;\ell_{\infty}(\mathcal{F})\quad
  $$
  where $\mathbb{G}_P$ is a tight Gaussian process.
  That is, it only remains to prove $\mathbb{G}_P=0$.
  This follows from considering finite marginals:
  Fix any $f_1,\dots,f_s\in\mathcal{F}$ and note that,
  due to (\ref{theorem-convergence-stochastic-term-p7})
  and (\ref{theorem-convergence-stochastic-term-p10}), the 
  (multivariate) Lindeberg-Feller central limit theorem
  \cite[e.g.,][Prop.\ 2.27]{vandervaart1998} for the random
  vectors 
  $$\big(g_{n,f_1}(Z_i,Y_i),\dots,g_{n,f_s}(Z_i,Y_i)\big),\qquad
    i\in\{1,\dots,n\},
  $$
  implies that $\big(\mathbb{G}_P(f_1),\dots,\mathbb{G}_P(f_s)\big)=0$.  
\end{proof}

\begin{proof}
  \item[\textbf{Proof of Prop.\ 
        \ref{prop-convergence-risk-deterministic-part}:}
       ]
  According to Assumption (\ref{theorem-unique-existence-theoretical-svm-1})
  the set $H_0:=\{f\in H\,|\,\mathcal{R}_{A,P}(f)<\infty\}$ is non-empty.
  Since 
  $B_f:[0,\infty)\rightarrow\mathbb{R},\;\;
   \lambda\mapsto\mathcal{R}_{A,P,\lambda}(f)
  $
  is continuous for every $f\in H_0$, the map
  $B:[0,\infty)\rightarrow\mathbb{R}$ defined by
  $B(\lambda)=\inf_{f\in H_0}B_f(\lambda)$, $\lambda\in[0,\infty)$,
  is upper semi-continuous.  For every $n\in\mathbb{N}$, 
  the function $f_{A,P,\lambda_n}\in H$ 
  uniquely exists according to Theorem 
  \ref{theorem-unique-existence-theoretical-svm} and
  we have 
  $B(\lambda_n)=\mathcal{R}_{A,P,\lambda_n}\big(f_{A,P,\lambda_n}\big)$.
  Then, the statement follows from
  \begin{eqnarray*}
    \lefteqn{
    \inf_{f\in H}\mathcal{R}_{A,P}(f)\;\leq\;
      \liminf_{n\rightarrow\infty}
          \mathcal{R}_{A,P}(f_{A,P,\lambda_n})
      \;\leq\;
      \limsup_{n\rightarrow\infty}
          \mathcal{R}_{A,P}(f_{A,P,\lambda_n})\;\leq
    }\\
    & &\leq\;\limsup_{n\rightarrow\infty} B(\lambda_n)
        \;\leq\;B(0)\;=\;\inf_{f\in H_0}\mathcal{R}_{A,P}(f)
        \;=\;\inf_{f\in H}\mathcal{R}_{A,P}(f)\qquad
  \end{eqnarray*}
\end{proof}

\begin{proof}
  \item[\textbf{Proof of Theorem 
        \ref{theorem-convergence-deterministic-part}:}
       ]
  First of all, it is shown that      
  \begin{eqnarray}\label{theorem-convergence-deterministic-part-p0}
    (f_n)_{n\in \mathbb{N}_0}\subset H,\;\;
    f_n\;\xrightarrow[]{\;\;\;\text{w}\;\;\;}\;f_0
    \quad\Rightarrow\quad
    \lim_{n\rightarrow\infty}\mathcal{R}_{A,P}(f_n)\,=\,\mathcal{R}_{A,P}(f_0).\quad
  \end{eqnarray}
  According to Prop.\ \ref{prop-compactness-A}, it follows from weak convergence
  of the sequence $(f_n)_{n\in \mathbb{N}_0}\subset H$ that
  $\lim_{n\rightarrow\infty}\big\|A(f_n)-A(f_0)\big\|_\infty=0$. Hence,
  there is an 
  $a\in(0,\infty)$ such that, for every $n\in\mathbb{N}_0$,
  we have $\big\|A(f_n)\big\|_\infty\leq a$.
  Then, (\ref{theorem-convergence-deterministic-part-p0}) follows from
  \begin{eqnarray*}
    \Big|\mathcal{R}_{A,P}(f_n)-\mathcal{R}_{A,P}(f_0)\Big|
    &\stackrel{(\ref{assumption-on-L-2})}{\leq}&
      \int \big(b_0^\prime+b_1^\prime a^p\big)\!\cdot\!\big|A(f_n)-A(f_0)|\,dP \\
    &\leq& \big\|A(f_n)-A(f_0)\big\|_{\infty}\!\cdot\!
           \int \big(b_0^\prime+b_1^\prime a^p\big)\,dP\,.\quad          
  \end{eqnarray*}
  Next, define 
  $$H_0\,:=\,
    \Big\{f^\ast\in H\,
    \Big|\,\mathcal{R}_{A,P}(f^\ast)=\inf_{f\in H}\mathcal{R}_{A,P}(f)
    \Big\}\;\stackrel{(\ref{theorem-convergence-deterministic-part-1})}{\not=}
    \;\emptyset\,.
  $$
  It follows from  
  (\ref{theorem-convergence-deterministic-part-p0}) that $H_0$ is closed in
  $H$. Furthermore, it follows from convexity of $t\mapsto L(z,y,t)$ that
  $H_0$ is a convex set. Hence, $H_0$ has a unique element $f_{A,P}$ of smallest
  norm; see, e.g., \cite[Theorem 3.7.9]{denkowski2003}. That is, we have shown
  unique existence of an element $f_{A,P}\in H$ which fulfills 
  (\ref{theorem-convergence-deterministic-part-2}) and 
  (\ref{theorem-convergence-deterministic-part-3}).\\
  The following proof of (\ref{theorem-convergence-deterministic-part-4})
  is similar to the proof of \cite[Theorem 5.17]{steinwart2008}.
  Let us fix
  a sequence $(\lambda_n)_{n\in\mathbb{N}}\subset(0,\infty)$ such that
  $\lim_{n\rightarrow\infty}\lambda_n=0$ and, first, prove
  \begin{eqnarray}\label{theorem-convergence-deterministic-part-p1}
    \big\|f_{A,P,\lambda_n}\big\|_H\;\leq\;\big\|f_{A,P}\big\|_H
    \qquad\forall\,n\in\mathbb{N}
  \end{eqnarray}
  by contradiction. If (\ref{theorem-convergence-deterministic-part-p1})
  was not true, then 
  $\big\|f_{A,P,\lambda_n}\big\|_H>\big\|f_{A,P}\big\|_H$ for some $n\in\mathbb{N}$
  and it would follow from $f_{A,P}\in H_0$ that 
  $\mathcal{R}_{A,P,\lambda_n}\big(f_{A,P,\lambda_n}\big)>
   \mathcal{R}_{A,P,\lambda_n}\big(f_{A,P}\big)
  $,
  which is a contradiction to the definition of $f_{A,P,\lambda_n}$.
  That is, we have shown (\ref{theorem-convergence-deterministic-part-p1}).
  Now, we prove (\ref{theorem-convergence-deterministic-part-4}) by contradiction.
  If $f_{A,P,\lambda_n}$ does not converge to $f_{A,P}$, then there is an
  $\varepsilon>0$ and a subsequence 
  $\big(f_{A,P,\lambda_{n_\ell}}\big)_{\ell\in\mathbb{N}}$ such that 
  \begin{eqnarray}\label{theorem-convergence-deterministic-part-p2}
    \big\|f_{A,P,\lambda_{n_\ell}}-f_{A,P}\big\|_H\;>\;\varepsilon
    \qquad\forall\,\ell\in\mathbb{N}.
  \end{eqnarray}
  Due to (\ref{theorem-convergence-deterministic-part-p1}), the subsequence
  $\big(f_{A,P,\lambda_{n_\ell}}\big)_{\ell\in\mathbb{N}}$ is bounded
  and, therefore, there is a further subsequence which converges weakly;
  see, e.g., \cite[Cor.\ IV.4.7]{dunford1958}.
  That is, we may assume without loss of generality that
  $\big(f_{A,P,\lambda_{n_\ell}}\big)_{\ell\in\mathbb{N}}$ fulfills
  (\ref{theorem-convergence-deterministic-part-p2}) and converges weakly to
  some $f_0\in H$. Then, it follows from 
  (\ref{theorem-convergence-deterministic-part-p0}) that 
  $\lim_{\ell\rightarrow\infty}\mathcal{R}_{A,P}(f_{A,P,\lambda_{n_\ell}})
   =\mathcal{R}_{A,P}(f_0)
  $. Hence, Prop.\ \ref{prop-convergence-risk-deterministic-part} implis
  that $\mathcal{R}_{A,P}(f_0)=\inf_{f\in H}\mathcal{R}_{A,P}(f)$ and,
  due to (\ref{theorem-convergence-deterministic-part-3}), we either have 
  $\|f_0\|_H > \|f_{A,P}\|_H$ or $f_0=f_{A,P}$. However,
  weak convergence implies
  $$\|f_0\|_H\;\leq\;
    \liminf_{\ell\rightarrow\infty}\big\|f_{A,P,\lambda_{n_\ell}}\big\|_H
    \;\leq\;\limsup_{\ell\rightarrow\infty}\big\|f_{A,P,\lambda_{n_\ell}}\big\|_H
    \;\stackrel{(\ref{theorem-convergence-deterministic-part-p1})}{\leq}\;
    \big\|f_{A,P}\big\|_H
  $$
  This proves $f_0=f_{A,P}$ and 
  $\lim_{\ell\rightarrow\infty}\|f_{A,P,\lambda_{n_\ell}}\|_H=\big\|f_{A,P}\big\|_H$.
  Finally, this latter convergence of the norms together with weak convergence of 
  $f_{A,P,\lambda_{n_\ell}}$ to $f_0=f_{A,P}$ implies
  $\lim_{\ell\rightarrow\infty}\big\|f_{A,P,\lambda_{n_\ell}}-f_{A,P}\big\|_H=0$;
  see, e.g., \cite[Exercise V.1.8]{conway1985}.
  This is a contradiction to (\ref{theorem-convergence-deterministic-part-p2}). 
\end{proof}

\begin{proof}
  \item[\textbf{Proof of Theorem 
        \ref{theorem-consistency}:}
       ]
  Theorem \ref{theorem-consistency} immediately follows from Theorem 
  \ref{theorem-convergence-stochastic-term} and
  \ref{theorem-convergence-deterministic-part}.
\end{proof}

\begin{proof}
  \item[\textbf{Proof of Theorem 
        \ref{theorem-rate-of-convergence-risks}:}
       ]
  Theorem \ref{theorem-consistency} guarantees 
  existence of $f_{A,P}\in H$ which fulfills  
  (\ref{theorem-consistency-2}) and (\ref{theorem-consistency-3}).
  As in the case of ordinary regularized kernel methods (i.e.\ $A=\text{id}$),
  \begin{eqnarray*}
    0&\leq&
         \mathcal{R}_{A,P}\big(f_{A,P,\lambda_n}\big)-\mathcal{R}_{A,P}^\ast\;\leq\;
         \mathcal{R}_{A,P,\lambda_n}\big(f_{A,P,\lambda_n}\big)-\mathcal{R}_{A,P}^\ast 
         \;\leq \\
     &\leq&\mathcal{R}_{A,P,\lambda_n}\big(f_{A,P}\big)-\mathcal{R}_{A,P}^\ast
         \;\stackrel{(\ref{theorem-consistency-2})}{=}\;
         \lambda_n\big\|f_{A,P}\big\|_H^2\;;
  \end{eqnarray*}
  see \cite[p.\ 182]{steinwart2008}. Hence,
  \begin{eqnarray}\label{theorem-rate-of-convergence-risks-p1}
    \Big|\mathcal{R}_{A,P}\big(f_{A,P,\lambda_n}\big)-\mathcal{R}_{A,P}^\ast
    \Big|
    \;\leq\;\lambda_n\big\|f_{A,P}\big\|_H^2\;.
  \end{eqnarray}
  Fix $n_0\in\mathbb{N}$ such that $a_n\geq 1$ for every $n\geq n_0$. 
  For every $\varepsilon\in(0,1)$, define
  $$B_{\varepsilon,n}\;:=\;
    \Big\{D_n\in(\mathcal{Z}\times\mathcal{Y})^n\,
    \Big|\,\,a_n\big\|f_{A,D_n,\lambda_n}-f_{A,P,\lambda_n}\big\|_\infty <\varepsilon
    \Big\}
    \quad\;\forall\,n\in\mathbb{N}
  $$
  and $a:=\sup_{n\in\mathbb{N}}\|f_{A,P,\lambda_n}\|_\infty+1$.
  It follows from Theorem \ref{theorem-convergence-deterministic-part}
  and (\ref{sup-norm-H-norm}) that $0<a<\infty$.
  Then, Assumption \ref{assumption-on-L} implies that, for every
  $D_n\in B_{\varepsilon,n}$ and $n\geq n_0$,
  \begin{eqnarray*}
    \lefteqn{
    a_n
    \Big|\mathcal{R}_{A,P}\big(f_{A,D_n,\lambda_n}\big)
         -\mathcal{R}_{A,P}\big(f_{A,P,\lambda_n}\big)
    \Big|
    \;\leq\;    
    }\\
    & &\leq
    \int \!b_0^\prime+b_1^\prime a^p\,dP \cdot
    a_n\big\|f_{A,D_n,\lambda_n}-f_{A,P,\lambda_n}\big\|_\infty
  \end{eqnarray*}   
  Since $\lim_{n\rightarrow\infty}P^n\big(B_{\varepsilon,n}\big)=1$
  for every $\varepsilon\in(0,1)$ according to
  Theorem \ref{theorem-convergence-deterministic-part}
  and (\ref{sup-norm-H-norm}), it follows that
  $$\lim_{n\rightarrow\infty}
    P\Big(\Big\{D_n\in(\mathcal{Z}\times\mathcal{Y})^n\,
          \Big|\,\,
               a_n
               \big|\mathcal{R}_{A,P}\big(f_{A,D_n,\lambda_n}\big)
                    -\mathcal{R}_{A,P}\big(f_{A,P,\lambda_n}\big)
               \big|
               <\varepsilon
          \Big\}
     \Big)
    =\,1
  $$
  for every $\varepsilon>0$. Then, (\ref{theorem-rate-of-convergence-risks-2})
  follows from (\ref{theorem-rate-of-convergence-risks-p1})
  and (\ref{theorem-rate-of-convergence-risks-1}).\\
  Finally, an elementary calculation
  shows that (\ref{theorem-rate-of-convergence-risks-1}) is fulfilled
  for $\lambda_n=\gamma n^{-1/(4+p)}\;\;\forall\,n\in\mathbb{N}$
  for a constant $\gamma\in(0,\infty)$.
\end{proof}

\begin{proof}
  \item[\textbf{Proof of Lemma 
        \ref{lemma-bounding-the-distance-between-functions-by-risks}:}
       ]
  According to the model assumptions, there is a set $B\in\mathfrak{B}_{\mathcal{Z}}$
  such that $P_{\mathcal{Z}}(B)=1$ and, for every $z\in B$,
  the conditional
  distribution $P(\cdot|z)$ is equal to the distribution of
  $t_z^\ast+s(z)\varepsilon$ where
  $t_z^\ast=\big(Af_{A,P}\big)(z)$ is the (unique) median of the conditional
  distribution $P(\cdot|z)$. Now fix any $z\in B$.
  Since $t_z^\ast$ is the median, 
  \begin{eqnarray}
    \label{lemma-bounding-the-distance-between-functions-by-risks-p4}
      P\big((-\infty,t_z^\ast]\,\big|\,z\big)
      \;\geq&\!\!\!\tfrac{1}{2}\!\!\!&\geq\;
      P\big((t_z^\ast,\infty)\,\big|\,z\big)\\
    \label{lemma-bounding-the-distance-between-functions-by-risks-p5}
      P\big([t_z^\ast,\infty)\,\big|\,z\big)
      \;\geq&\!\!\!\tfrac{1}{2}\!\!\!&\geq\;
      P\big((-\infty,t_z^\ast)\,\big|\,z\big)
  \end{eqnarray} 
  and the model assumptions 
  (\ref{heteroscedastic-regression-model-nochmal})--%
  (\ref{heteroscedastic-regression-model-assumption-error-density-2})
  imply
  \begin{eqnarray}
    \label{lemma-bounding-the-distance-between-functions-by-risks-p6}
    P\Big(\big(t_z^\ast\,,\,t_z^\ast+\delta\big)\,\Big|\,z\Big)
    &\geq&\;\;\frac{c_h}{\overline{c}_s}\cdot\delta \qquad\quad
        \forall\,\delta\in(0,\alpha) \\
    \label{lemma-bounding-the-distance-between-functions-by-risks-p7}
    P\Big(\big(t_z^\ast-\delta\,,\,t_z^\ast\big)\,\Big|\,z\Big)
    &\geq&\;\;\frac{c_h}{\overline{c}_s}\cdot\delta \qquad\quad
        \forall\,\delta\in(0,\alpha)\,. 
  \end{eqnarray} 
  The rest of the proof is divided into different cases;
  first, we consider the case that
  $t>t_z^\ast$. Note that
  \begin{eqnarray}
    \label{lemma-bounding-the-distance-between-functions-by-risks-p2}
    |y-t|-|y-t_z^\ast|
    &>&\;\;0 \qquad\quad\;
        \text{if }\;\;t_z^\ast\,<\,y\,<\,\tfrac{1}{2}(t_z^\ast+t)\qquad\qquad \\
    \label{lemma-bounding-the-distance-between-functions-by-risks-p3}
    |y-t|-|y-t_z^\ast|
    &>&t_z^\ast-t 
       \qquad\text{if }\;\;\tfrac{1}{2}(t_z^\ast+t)\,\leq\,y\,<\,t\qquad\qquad 
  \end{eqnarray} 
  Then, dividing the domain of integration 
  by $t_z^\ast$, $\tfrac{1}{2}(t_z^\ast+t)$, and $t$ 
  into four parts yields
  \begin{eqnarray*}
    \lefteqn{
    \int\! L(z,y,t)\,P(dy|z)\,-\int\! L(z,y,t_z^\ast)\,P(dy|z)
    \;=\int |y-t|\,-\,|y-t_z^\ast|\,P(dy|z) 
    }\\
    &\stackrel{(\ref{lemma-bounding-the-distance-between-functions-by-risks-p2}),
               (\ref{lemma-bounding-the-distance-between-functions-by-risks-p3})
              }{\geq}& 
         (t-t_z^\ast)\cdot P\Big(\big(-\infty,t_z^\ast\big]\,\Big|\,z\Big)\,+\,
           0\cdot 
             P\Big(\big(t_z^\ast\,,\,\tfrac{1}{2}(t_z^\ast+t)\big)\,\Big|\,z\Big)
           - \\
         &&\;\; -\;(t-t_z^\ast)\cdot
             P\Big(\big(\tfrac{1}{2}(t_z^\ast+t)\,,\,t\big)\,\Big|\,z\Big)
             \,-\,
			(t-t_z^\ast)\cdot P\Big(\big[t,\infty\big)\,\Big|\,z\Big)\quad\;\\
    &\stackrel{(\ref{lemma-bounding-the-distance-between-functions-by-risks-p4})}{\geq}&
          (t-t_z^\ast)\cdot
          P\Big(\big(t_z^\ast\,,\,\tfrac{1}{2}(t_z^\ast+t)\big)\,\Big|\,z\Big)
        \;\stackrel{(\ref{lemma-bounding-the-distance-between-functions-by-risks-p6})
                   }{\geq}\;
                \frac{c_h}{2\overline{c}_s}(t-t_z^\ast)^2\,.
  \end{eqnarray*}
  Next, we consider the case that
  $t<t_z^\ast$. Similarly as before,
  dividing the domain of integration 
  by $t$, $\tfrac{1}{2}(t_z^\ast+t)$, and $t_z^\ast$ 
  into four parts yields
  \begin{eqnarray*}
    \lefteqn{
    \int\! L(z,y,t)\,P(dy|z)\,-\int\! L(z,y,t_z^\ast)\,P(dy|z)
    \;=\int |y-t|\,-\,|y-t_z^\ast|\,P(dy|z) 
    }\\
    &\geq& 
         -(t_z^\ast-t)\cdot P\Big(\big(-\infty,t\big]\,\Big|\,z\Big)\,-\,
            (t_z^\ast-t)\cdot 
             P\Big(\big(t\,,\,\tfrac{1}{2}(t_z^\ast+t)\big]\,\Big|\,z\Big)
           - \\
         &&\;\; +\;0\cdot
             P\Big(\big(\tfrac{1}{2}(t_z^\ast+t)\,,\,t_z^\ast\big)\,\Big|\,z\Big)
             \,+\,
			(t_z^\ast-t)\cdot P\Big(\big[t_z^\ast,\infty\big)\,\Big|\,z\Big)\quad\;\\
    &\stackrel{(\ref{lemma-bounding-the-distance-between-functions-by-risks-p5})}{\geq}&
          (t_z^\ast-t)\cdot
          P\Big(\big(\tfrac{1}{2}(t_z^\ast+t)\,,\,t_z^\ast\big)\,\Big|\,z\Big)
        \;\stackrel{(\ref{lemma-bounding-the-distance-between-functions-by-risks-p7})
                   }{\geq}\;
                \frac{c_h}{2\overline{c}_s}(t_z^\ast-t)^2\,.
  \end{eqnarray*}
  Finally, the remaining case $t=t_z^\ast$ is trivial.
\end{proof}

\begin{proof}
  \item[\textbf{Proof of Theorem 
        \ref{theorem-rate-of-convergence}:}
       ]
  First, note that $P$ and $L$ fulfill
  Assumption \ref{assumption-on-L} for $p=0$ and that the model assumptions
  (\ref{heteroscedastic-regression-model-nochmal})--%
  (\ref{heteroscedastic-regression-model-assumption-error-density-2})  
  imply $\mathcal{R}_{A,P}(f_{A,P})=\inf_{f\in H}\mathcal{R}_{A,P}(f)$.\\
  According to (\ref{assumption-operator-A-continuous-on-H}) and
  Theorem \ref{theorem-convergence-deterministic-part}, 
  there is an $n_0\in\mathbb{N}$ such that
  $\|Af_{A,P,\lambda_n}-Af_{A,P}\|_\infty\leq \|A\|
   \cdot\|f_{A,P,\lambda_n}-f_{A,P}\|_H < a_h\underline{c}_s=:\alpha
  $
  for every $n\geq n_0$. Hence, it follows from Lemma
  \ref{lemma-bounding-the-distance-between-functions-by-risks} 
  and (\ref{theorem-rate-of-convergence-risks-p1}) in the proof
  of Theorem \ref{theorem-rate-of-convergence-risks} 
  that
  there is a constant $\tilde{c}\in(0,\infty)$ such that 
  \begin{eqnarray}\label{theorem-rate-of-convergence-p1}
    \big\|Af_{A,P,\lambda_n}-Af_{A,P}\big\|_{L_2(P_{\mathcal{Z}_0})}
    \;\leq\;\tilde{c}\lambda_n^\frac{1}{2}
    \qquad\forall\,n\geq n_0\,.
  \end{eqnarray}
  According to Assumption (\ref{theorem-rate-of-convergence-2}),
  $$g_0\;:=\;
    \sum_{j=1}^\infty 
         \frac{\big\langle f_{A,P},v_j\big\rangle_H}{\sigma_j}\cdot u_j
    \;\;\in\;\;L_2(P_{\mathcal{Z}_0})
  $$
  and it follows from (\ref{singular-value-decomposition-1}) and the fact that
  $\{v_j\,|\,j\in\mathbb{N}\}$ is a complete orthonormal system
  of $H$ that
  \begin{eqnarray}\label{theorem-rate-of-convergence-p2}
    f_{A,P}=A_0^\ast g_0\,.
  \end{eqnarray}
  An easy calculation shows
  \begin{eqnarray}\label{theorem-rate-of-convergence-p3}
    \big\|f_{A,P,\lambda_n}\!-\!f_{A,P}\big\|_H^2
       \!-\!\big\|f_{A,P,\lambda_n}\big\|_H^2
    = 2\big\langle f_{A,P}\!-\!f_{A,P,\lambda_n},f_{A,P}\big\rangle_{\!H}
      \!-\!
      \big\|f_{A,P}\big\|_H^2.\;
  \end{eqnarray}
  According to the definitions, we have
  \begin{eqnarray}
    \label{theorem-rate-of-convergence-p4}
    \mathcal{R}_{A,P}\big(f_{A,P,\lambda_n}\big)-
    \mathcal{R}_{A,P}\big(f_{A,P}\big)
    &\geq&0\qquad\forall\,n\in\mathbb{N} \\
    \label{theorem-rate-of-convergence-p5}
    \mathcal{R}_{A,P,\lambda_n}\big(f_{A,P,\lambda_n}\big)-
    \mathcal{R}_{A,P,\lambda_n}\big(f_{A,P}\big)
    &\leq&0\qquad\forall\,n\in\mathbb{N}\,.\qquad
  \end{eqnarray}
  Recall that 
  $\mathcal{R}_{A,P,\lambda_n}(f)=\mathcal{R}_{A,P}(f)+\lambda_n\|f\|_H^2$.
  Hence,
  \begin{eqnarray*}
    \lefteqn{
    \lambda_n\big\|f_{A,P,\lambda_n}\!-\!f_{A,P}\big\|_H^2
    \;\leq
    }\\
    &\!\stackrel{(\ref{theorem-rate-of-convergence-p4})}{\leq}&
       \!\!\!
       \mathcal{R}_{A,P}\big(f_{A,P,\lambda_n}\big)-
         \mathcal{R}_{A,P}\big(f_{A,P}\big)
       +\lambda_n\big\|f_{A,P,\lambda_n}\big\|_H^2 - \\
     &&\;\;+\;\lambda_n\big\|f_{A,P,\lambda_n}\!-\!f_{A,P}\big\|_H^2
              -\lambda_n\big\|f_{A,P,\lambda_n}\big\|_H^2\;=\\
    &\!\stackrel{(\ref{theorem-rate-of-convergence-p3})}{=}&
       \!\!\!
       \mathcal{R}_{A,P,\lambda_n}\big(f_{A,P,\lambda_n}\big)\!-\!
       \mathcal{R}_{A,P,\lambda_n}\big(f_{A,P}\big)
       +2\lambda_n\big\langle f_{A,P}\!-\!f_{A,P,\lambda_n},f_{A,P}\big\rangle_{\!H}
       \leq\\
    &\!\stackrel{(\ref{theorem-rate-of-convergence-p5}),
                 (\ref{theorem-rate-of-convergence-p2})
                }{\leq}&
       \!\!\!
       2\lambda_n\big\langle f_{A,P}\!-\!f_{A,P,\lambda_n},A_0^\ast g_0\big\rangle_{\!H}
       \,=\,2\lambda_n\big\langle A_0f_{A,P}\!-\!A_0f_{A,P,\lambda_n},
                                  g_0
                      \big\rangle_{L_2(P_{\mathcal{Z}_0})} \\
    &\!\leq&\!\!\!
       2\lambda_n
       \big\|A_0f_{A,P,\lambda_n}\!-\!A_0f_{A,P}\big\|_{L_2(P_{\mathcal{Z}_0})}
       \cdot\big\|g_0\big\|_{L_2(P_{\mathcal{Z}_0})}.                         
  \end{eqnarray*}
  Together with (\ref{theorem-rate-of-convergence-p1}), this implies
  that there is a constant $c\in(0,\infty)$ such that
  $$\big\|f_{A,P,\lambda_n}\!-\!f_{A,P}\big\|_H^2
    \;\leq\;c\lambda_n^\frac{1}{2}
    \qquad\forall\,n\geq n_0\,.
  $$
  Hence, if $\lim_{n\rightarrow\infty}a_n\lambda_n^{\frac{1}{2}}=0$,
  it follows from Theorem \ref{theorem-convergence-stochastic-term} that
  $$a_n\big\|f_{A,\mathbf{D}_n,\lambda_n}-f_{A,P}\big\|_H^2
    \;\;\xrightarrow[\;n\rightarrow\infty\;]{}\;\;0
    \qquad\text{in probability}.
  $$
  Next, let Assumption (\ref{theorem-rate-of-convergence-7}) be fulfilled. 
  Then,
  $$g\;:=\;\sum_{j=1}^\infty \frac{v_j(x)}{\sigma_j}\cdot u_j\;\;\in\;\;
           L_2\big(P_{\mathcal{Z}_0}\big)
  $$
  and 
  $$A_0^\ast g
    \;\stackrel{(\ref{singular-value-decomposition-1})}{=}\;
      \sum_{j=1}^\infty v_j(x)\cdot v_j
    \;\stackrel{(\ref{reproducing-property})}{=}\;
       \sum_{j=1}^\infty \langle v_j,\Phi(x)\rangle_{H}\cdot v_j
       \;=\;\Phi(x)
  $$
  where we have used in the last equality that 
  $\{v_j\,|\,j\in\mathbb{N}\}$ is a complete orthonormal system
  of $H$. Hence,
  \begin{eqnarray*}
    \lefteqn{
    f_{A,P,\lambda_n}(x)-f_{A,P}(x)
    \;\stackrel{(\ref{reproducing-property})}{=}\;
       \big\langle f_{A,P,\lambda_n}-f_{A,P},\Phi(x) \big\rangle_H \;=
    }\\
    &=&\big\langle f_{A,P,\lambda_n}-f_{A,P}, A_0^\ast g \big\rangle_H
       \;=\;\big\langle A_0 f_{A,P,\lambda_n}-A_0f_{A,P}\,,\, g 
            \big\rangle_{L_2(P_{\mathcal{Z}_0})}
  \end{eqnarray*}
  and, accordingly, it follows from 
  (\ref{theorem-rate-of-convergence-p1}) that there is a constant
  $c\in(0,\infty)$ such that
  $$\big(f_{A,P,\lambda_n}(x)-f_{A,P}(x)\big)^2\;\leq\;
    c\lambda_n \qquad \forall\,n\geq n_0\,.
  $$
  Hence, if $\lim_{n\rightarrow\infty}a_n\lambda_n=0$,
  it follows from Theorem \ref{theorem-convergence-stochastic-term} 
  and the reproducing property (\ref{reproducing-property})
  that
  $$a_n\big(f_{A,\mathbf{D}_n,\lambda_n}(x)-f_{A,P}(x)\big)^2
    \;\;\xrightarrow[\;n\rightarrow\infty\;]{}\;\;0
    \qquad\text{in probability}.
  $$
\end{proof}

\bibliographystyle{abbrvnat}
\bibliography{literatur}

\end{document}